\documentclass[aps,prx,twocolumn,superscriptaddress,nobalancelastpage,amsmath,amssymb,nofootinbib,longbibliography]{revtex4-1}
\usepackage[utf8]{inputenc}
\usepackage{float}
\usepackage{amsmath}
\usepackage{tikz}
\usepackage{graphicx}% Include figure files
\usepackage{dcolumn}% Align table columns on decimal point
\usepackage{multirow}

\def\[#1\]{%
  \begin{equation}#1\end{equation}%
}

\usepackage{framed, ulem}
\definecolor{shadecolor}{gray}{0.95}

\usepackage{bm}% bold math
\newcommand{\bZ}{\mathbb{Z}}

\newcommand{\cO}{\mathcal{O}}
\usepackage{hyperref}
\hypersetup{
    pdftitle={Gapless SPT},
    unicode=false,          % non-Latin characters 
    pdftoolbar=true,        % show Acrobat
    pdfmenubar=true,        % show Acrobat 
    pdffitwindow=false,     % window fit to page when opened
    pdfstartview={FitH},    % fits the width of the page to the window
    pdfsubject={},   % subject of the document
    pdfcreator={},   % creator of the document
    pdfproducer={}, % producer of the document
    pdfkeywords={} {} {}, % list of keywords
    pdfnewwindow=true,      % links in new window
    colorlinks=true,       % false: boxed links; true: colored links
    linkcolor=blue, %red,          % color of internal links (change box color with linkbordercolor)
    citecolor=blue,        % color of links to bibliography
    filecolor=blue,      % color of file links
    urlcolor=blue           % color of external links
}

\begin{document}
%\title{Emergent Anomalies and Exotic Topological Criticality in the Spin-Gapped Hubbard Chain}
%\title{Emergent Anomalies and Exotic Topological Criticality}
%\title{Topological criticality protected by emergent anomalies}

%\title{Not all gapless topological phases are separable into gapped topological phases and conventional criticality}

%\title{Gapless topological phases without gapped counterpart: anomalies with an edge in the Ising-Hubbard chain}
\title{Intrinsically Gapless Topological Phases}%: anomalies in the Ising-Hubbard chain}

\author{Ryan Thorngren}
\affiliation{Center for Mathematical Sciences and Applications, Harvard University, Cambridge, MA 02138, USA}

\author{Ashvin Vishwanath}
\affiliation{Department of Physics, Harvard University, Cambridge, MA 02138, USA}

\author{Ruben Verresen}
\affiliation{Department of Physics, Harvard University, Cambridge, MA 02138, USA}

\begin{abstract}
%Topological order is typically defined for gapped phases, where the finite correlation length localizes protected edge modes at the boundary and protects superselection sectors in the bulk. Recently, it has been shown that a particular class, namely symmetry-protected topological (SPT) phases, can be stable to coupling to gapless degrees of freedom. In this paper, however, we explore topological phases for which the vanishing of the energy gap is essential, and there is no gapped counterpart. We call such topological phases intrinsically gapless. A natural one-dimensional example is given by the Ising-Hubbard chain, which we analyze in detail. The topology of the nontrivial gapless phase of this model reveals itself as a zero energy edge mode protected by a $\bZ_4$ global on-site symmetry, whereas there are no gapped SPTs in 1D protected by this group. We propose emergent anomalies as a general mechanism for obtaining intrinsically gapless SPT phases, such as in the Ising-Hubbard chain, where the microscopic $\bZ_4$ effectively acts as an anomalous $\bZ_2$ at low energies. Such anomalies can be diagnosed using string order parameters, from which one can infer the existence of protected zero-energy edge modes. More generally, emergent anomalies for on-site symmetries always imply edge modes.
Topology in quantum matter is typically associated with gapped phases. For example, in symmetry protected topological (SPT) phases, the bulk energy gap localizes edge modes near the boundary. In this work we identify a new mechanism that leads to topological phases which are not only gapless but where the absence of a gap is essential. These `intrinsically gapless SPT phases' have no gapped counterpart and are hence also distinct from recently discovered examples of gapless SPT phases. The essential ingredient of these phases is that on-site symmetries act in an anomalous fashion at low energies. Intrinsically gapless SPT phases are found to display several unique properties including (i) protected edge modes that are impossible to realize in a gapped system with the same symmetries, (ii) string order parameters that are likewise forbidden in gapped phases, and (iii) constraints on the phase diagram obtained upon perturbing the phase. We verify predictions of the general theory in a specific realization protected by $\mathbb Z_4$ symmetry, the one dimensional Ising-Hubbard chain, using both numerical simulations and effective field theory. We also discuss extensions to higher dimensions and possible experimental realizations.
\end{abstract}

\date{\today}
\maketitle

\section{Introduction}

Topology appears in various fascinating guises in many-body quantum physics.
%In this letter, we focus on the broad class of 
For instance, symmetry-protected topological (SPT) phases  generalize the notion of topological insulators and superconductors \cite{Hasan10} to include interactions and to other symmetry classes \cite{WenRMP,Senthil_2015}. Such systems are well understood when there is a gapped symmetric bulk protecting the edge modes.
%Many recent works have addressed the question of the fate of these edge modes when the bulk gap closes \cite{}.
%There has been considerable study in recent years on characterizing when these edge modes can persist when the bulk gap closes \cite{}.
%In recent years, there have been a multitude of works showing that such edge modes can sometimes persist when the bulk gap closes \cite{}.

%One question is whether there are fundamentally new topological phenomena occurring in gapless systems.

Previous work has also established
%focused on 
the remarkable stability of gapped SPT physics upon closing the bulk gap \cite{Kestner11,Fidkowski11,Sau11,Tsvelik11,Cheng11,Ruhman12,Grover12,Kraus13,Ortiz14,Ruhman15,Iemini15,Lang15,Keselman15,Kainaris15,Ortiz16,Montorsi17,Wang17,Kane17,Ruhman17,Scaffidi_2017,Guther17,Kainaris17,Chen18,Verresen18,Zhang18,Jiang18,Parker18,Keselman18,Jones19,verresen_gapless_2019,verresen2020topology}.
%and does not address this question. 
Here we will instead be interested in a different question---can fundamentally \textit{new} topological phenomena occur in gapless systems?
%In this work, however we show that the world of gapless topological phases is indeed quite rich: there exist gapless SPT phases which in a precise sense have no gapped counterpart.
In this letter  
%, however, 
we show that gapless topological phases considerably expand our notion of what is possible: there exist gapless SPT phases which in a precise sense have no gapped counterpart. For concreteness, in this work we will focus on such examples in 1+1D, although we briefly discuss how the physics can be generalized to higher dimensions.

These intrinsically gapless SPT phases share 
certain properties with gapped SPT phases. They have an on-site symmetry action and their topology can be diagnosed---at least in 1+1D---by studying the charges residing at the ends of string order parameters. Furthermore, the nontrivial string charges imply the existence of exponentially localized zero-energy edge modes. However, while in a gapped phase such charges are constrained to take certain values associated with projective representations of the symmetry group \cite{Fidkowski_2011,pollmann_symmetry_2012,chen_complete_2011,Schuch11}, in these gapless phases there is no such restriction. In fact we will see the charge assignments for intrinsically gapless SPT string order parameters are forbidden in gapped phases.
%interpretation, and the charges may take more general values.

As an illustrative example, we show that charge doping an Ising chain restores spin flip symmetry in the bulk but leaves it broken on the boundary, leading to edge modes. This is similar to an SPT phase but differs in at least three respects. First, the bulk contains a Luttinger liquid of gapless charge fluctuations. Second, any small perturbation which opens a gap will restore antiferromagnetic order, so that the topological edge modes only occur in the gapless region. Third, the system has a long-range ordered string operator whose charge is forbidden in any gapped phase.
While earlier studies of related gapless models \cite{Kruis_2004,Keselman15,Kainaris15,Kainaris17} observed string order and edge modes, we focus on the underlying mechanism which forms a definite break from gapped SPT physics.

%While earlier studies of this and related gapless models \cite{Kruis_2004} noticed an unusual form of hidden order, the connection to topological phases was missing. In particular the edge modes identified here and the emergent anomaly we discuss shortly were not considered before. \av{Comment: Should we cite all the other relevant papers here briefly - Refs [Berg, Carr..] noted other symmetry settings which admit gapless SPTs which have some overlap with the current work, although the emergent anomaly which we discuss shortly was not considered before.}

% In 1d, SPT order can be diagnosed by the study of string order parameters \cite{pollmann_symmetry_2012}. \av{I would move the next sentence to the gapless SPT paragraph} In Ref.~\cite{verresen_gapless_2019} we showed that this concept extends just as well to gapless systems. (We review these developments in Appendix \ref{appbiganomstring}). \av{String order parameters of conventional SPTs are never charged under their own symmetry. in contrast,..} Remarkably, for our topological Luttinger liquid we find that the string order parameters have charges which are inconsistent with a gapped bulk, demonstrating that the phase is indeed intrinsically gapless.

We propose a general mechanism to understand such nontrivial string order parameters, the protected edge modes, and the curious nearby phase diagrams in terms of emergent anomalies. Anomalies yield powerful constraints on the ground states of systems. Indeed, they play a  key role in determining the gapless/symmetry breaking behavior of our 1+1D model. However, anomalies are  usually associated with the edge of a topological bulk, for example the 1+1D edge of a 2+1D nontrivial SPT which must exhibit either gapless behavior or broken symmetry \cite{Qi_2008,Levin_2012,Chen_2013,Else_2014,Witten_2016}. How can anomalies can be relevant for a genuine 1+1D system? There are two ways. The first is if the symmetry action cannot be characterized as `on-site'. This arises for example in the spin 1/2 chain, where  translation symmetry, which is clearly not an on-site symmetry,  leads to an anomaly and there are associated Lieb-Schultz-Mattis theorems which constrain the phases of this system \cite{Lieb61,Cho_2017,Jian_2018,Metlitski_2018,Cheng2016, Yang_2018}. The Fermi-arc surface states \cite{Wan2011}  of Weyl semi-metals can similarly be interpreted as `impossible' surface states arising from an anomalous 3+1 D bulk. However in that case as well, translation symmetry is invoked.
%(e.g., translation symmetry is anomalous in spin-$1/2$ chains). 
The second way---explored in this work since our symmetries are on-site---is that of an {\bf emergent} anomaly. %which leads to  but they tend to occur for systems with a non-locality: either they live at the boundary of a higher-dimensional system, such as an anomalous SPT edge \cite{Qi_2008,Levin_2012,Chen_2013,Else_2014,Witten_2016}, or their symmetry action does not obey a tensor product structure, such as with translation in Lieb-Schultz-Mattis theorems \cite{Cho_2017,Jian_2018,Metlitski_2018,Yang_2018}.

An emergent anomaly can occur whenever the (non-anomalous) microscopic symmetry is not faithfully represented on the gapless modes, due to fundamental charges being gapped \cite{Metlitski_2018}. In the doped Ising chain example that we shall discuss, electrons are gapped, so fermion parity symmetry acts trivially on all gapless modes. 
The effective low-energy symmetry is the quotient of the full symmetry group by the part that only acts on gapped modes.
The emergent anomaly of the low-energy symmetry can be diagnosed from the charges of the string order parameters, which is nontrivial whenever these are inconsistent with any gapped SPT.

A fundamentally new feature of emergent anomalies arises when the microscopic symmetry is \textit{on-site}, since one can then consider boundaries for which the symmetry is well-defined.
%When the anomaly occurs for an on-site symmetry, as in our examples, it makes sense to consider symmetric boundary conditions.
Usually, since anomalies live on the edge of a higher-dimensional system, this is tantamount to  considering
%This is a bit like considering
the ``boundary of a boundary" which goes beyond the scope of the conventional theory of anomalies. 
%is not described
%goes beyond by the existing theory of anomalies.
%We give a partial characterization using the string operators, and argue 
We argue that emergent anomalies of such on-site symmetries lead to nontrivial edge modes protected by symmetry and gaplessness.

%This also gives a new perspective on 2+1D bosonic SPT phases. It has been observed that these can sometimes be trivialized by embedding them into a bigger Hilbert space---a natural instance being the addition of fermions \cite{Lu12,Lu14,Wang14,Gu14,Wang_2016}---meaning that the edge theory is no longer absolutely protected. The present work shows that there is still a well-defined emergent anomaly at the edge, and it can thus be realized in its own dimension as an intrinsically gapless topological phase.

%The emergent anomalies we study in this paper are associated to 2+1D bosonic SPTs which become trivial in the presence of fundamental fermions. Thus, although such phases cannot be made from electron systems, their anomalous edge theories can exist on their own as an intrinsically gapless topological phase.

The remainder of this paper is structured as follows. In Section~\ref{seclattice} we introduce the Ising-Hubbard chain which gives a simple realization of an intrinsically gapless SPT phase whose edge modes and unusual string order are protected by $\mathbb Z_4$ symmetry. The emergent anomaly of this topological phase is more readily apparent in the field theoretic perspective explored in Section~\ref{secfieldtheory}. The key ingredients are summarized and generalized in Section~\ref{secgenframe}, applicable to any dimension and symmetry group $G$.

\begin{figure}
    \centering
    \includegraphics[scale=1]{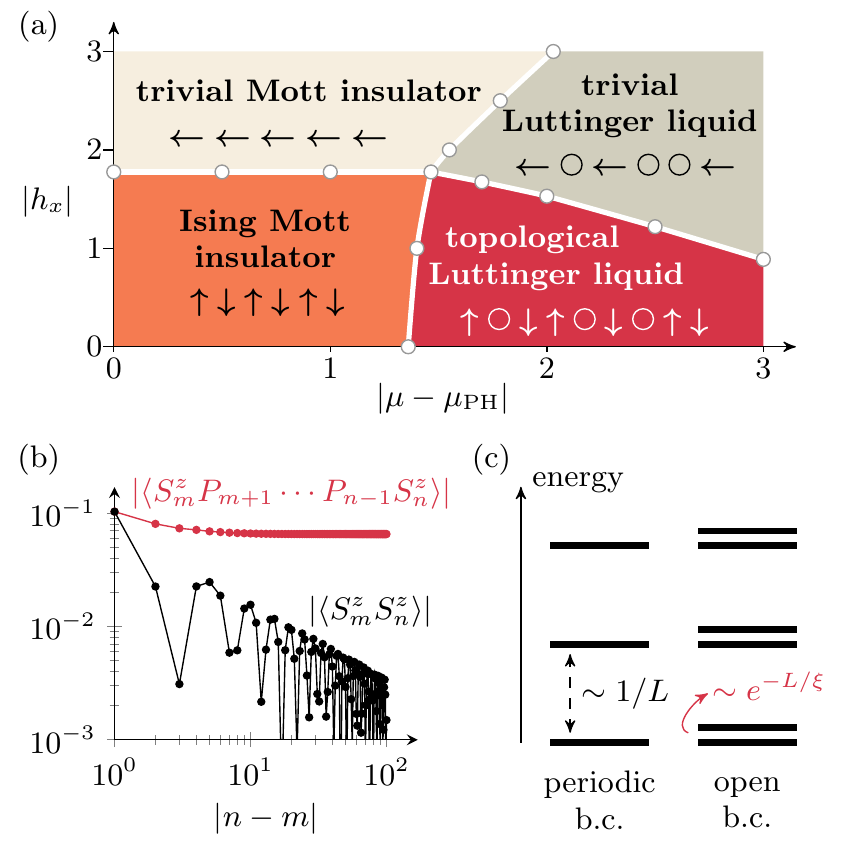}
    \caption{\textbf{Intrinsically gapless SPT phase in the Ising-Hubbard chain.} (a) Phase diagram for the model in Eqs.~\eqref{eq:Ising} and \eqref{eq:Hub} for $t=J_z=1$ and $U=5$. The chemical potential is relative to the particle-hole-symmetric value $\mu_\textrm{PH}=U/2$. The trivial and symmetry-breaking phases for small chemical potential are in the Mott limit. Eventually, these make way for a trivial and topological Luttinger liquid (LL), respectively. The topology of the latter is protected by $\mathbb Z_4$; if one additionally imposes translation symmetry, this is a stable gapless phase.
    (b) The topological order (or hidden symmetry breaking) in the topological LL can be detected by the string order parameter (plotted for $|\mu - \mu_\textrm{PH}|=2$ and $h_x = 0$).
    (c) Whilst the topological LL has a unique ground state for periodic boundary conditions (b.c.), it has a twofold degeneracy for open b.c. whose exponentially-small splitting is determined by the spin correlation length.}
    \label{fig:dopedIsing}
\end{figure}

\section{An Exotic Gapless Topological Phase}\label{seclattice}

Let us describe a simple model which illustrates this phenomenon, namely the doped Ising model, or equivalently, 
the Ising-Hubbard model.
%the Hubbard model with an Ising coupling.
This describes a chain of spinful fermions $c^{\dagger}_s$ with Hamiltonian $H = H_\textrm{Ising}+ H_\textrm{Hub} $, where
\begin{align}
H_\textrm{Ising} &= \sum_n \left( J_z S^z_{n} S^z_{n+1} + h_x S^x \right) \label{eq:Ising}\\
H_\textrm{Hub} &= -t \sum_{j,s} \left(c^\dagger_{j+1,s} c_{j,s}^{\vphantom \dagger} + h.c.\right) + U \sum_j n_{j,\uparrow} n_{j,\downarrow} - \mu N \label{eq:Hub}
\end{align}
%\[\label{eqnhubbJz} H = \sum_{j,s} \left(c^\dagger_{j,s} c_{j,s}^{\vphantom \dagger} + h.c.\right) + \sum_j \left( U n_{j,\uparrow} n_{j,\downarrow} + J_z S^z_j S^z_{j+1} \right),\]
with $n_{j,s} = c^\dagger_{j,s} c_{j,s}^{\vphantom \dagger} $ and $S^\alpha_j = \frac{1}{2} c^\dagger_{j,s} \sigma^\alpha_{s,s'} c_{j,s'}^{\vphantom \dagger} $. %Crucially, spin rotation is partially broken by the $J_z$ term to the group generated by $z$-axis rotations and a
The spin rotation is broken by the $J_z$ and $h_x$ terms to the $\pi$-rotation $R_x$ around the $x$-axis. Note that this is a $\mathbb Z_4$ symmetry since $R_x^2 = (-1)^F = P$.
At half-filling, the Hubbard interaction $U$ drives the model into a Mott phase (i.e., $\langle n_j \rangle = 1$), such that we obtain an effective spin-$1/2$ chain which is either in an Ising phase (spontaneously breaking $R_x$ down to its $\mathbb Z_2$ subgroup of fermion parity) or a symmetry-preserving paramagnet, depending on the value of $h_x$. However, if we dope the system using a chemical potential (i.e., $\langle n_j\rangle \neq 1$), a gapless Luttinger liquid appears, while spins remain gapped. The numerical phase diagram for $t=J_z=1$ and $U=5$ is shown in Fig.~\ref{fig:dopedIsing}(a), which was obtained using the density matrix renormalization group (DMRG) \cite{White93,Hauschild18}.

The key point is that doping the Ising antiferromagnet means that we lose long-range order in $S^z$ since an arbitrary number of holes (or doublons) can appear between any two antiferromagnetically-aligned spins. However, as sketched in Fig.~\ref{fig:dopedIsing}(a), the resulting Luttinger liquid retains a hidden symmetry-breaking pattern where the ``squeezed state" with the holes (or doublons) removed retains anti-ferromagnetic correlations \cite{Zhang97,Kruis_2004}. This can be probed by measuring the $S^z$-correlation function whilst keeping tracking of all the holes, which is done by inserting a string of fermionic parity: $\langle S^z_m \big( \prod_{m<k<n}P_k \big) S^z_n \rangle$. Figure~\ref{fig:dopedIsing}(b) shows that this indeed has long-range order in the Luttinger liquid adjacent to the Ising phase, while the local Ising order parameter decays algebraically.
%\av{maybe reiterate what was known and what's not in reference 13, for example now we discuss string order parameters and why they cannot be SPT like and also the resulting edge modes that were not part of earlier studies.}

While this string order parameter was pointed out before \cite{Zhang97,Kruis_2004,Rakovszky20}, it was not appreciated how this implies that the ground state is topologically non-trivial with respect to $\mathbb Z_4$. As we now discuss, (i) this implies the presence of edge modes and (ii) a gapped SPT phase can never host this particular string order---hence this constitutes an intrinsically gapless topological state.
%implying edge modes, as we discuss now. In fact, no gapped SPT phase can host this particular string order parameter.

The $R_x$ symmetry is fractionalized on this order parameter in the sense that its endpoints, which carry an $S^z$ insertion, are charged. This is a hallmark feature of symmetry-protected topological (SPT) phases \cite{pollmann_symmetry_2012}, harking back to the seminal work by den Nijs and Rommelse \cite{dennijs89}. Usually, SPT phases are gapped, yet here we have long-range order in a string order parameter despite the system being gapless. Such gapless topological phases have been studied before \cite{Kestner11,Cheng11,Grover12,Kraus13,Ortiz14,Iemini15,Lang15,Keselman15,Kainaris15,Montorsi17,Wang17,Scaffidi_2017,Guther17,Chen18,Verresen18,Zhang18,Jiang18,Parker18,Keselman18,Jones19,verresen_gapless_2019}. However, what is entirely novel is that in our example, the topological phase is protected by $\mathbb Z_4$ alone;
%\footnote{With just $R_x$ symmetry, the topological phase has a pair of relevant deformations into a gapped antiferromagnet. We describe the phase diagram in Appendix \ref{appnearbyphasediagram}. We note that $\mathbb Z_4 \times \mathbb Z_3$ stabilizes the gapless topological phase---but still does not admit any gapped SPT.};
indeed, the charge of the above string operator\footnote{We automatically have $P$ since it is an unbreakable fermionic symmetry. Alternatively, $R_x^2 = P$---hence the same physics occurs in purely bosonic system with $\bZ_4$ symmetry, see Appendix~\ref{appothersymmetryclasses}.} is well-defined as long as we preserve $R_x$.
This is the first example that goes beyond the gapped classification (since $H^2(\mathbb Z_4,U(1)) =0$\footnote{Strictly speaking, for fermions these phases are classified by spin cobordism. The group is still zero, however \cite{hason2019anomaly}.}), giving an intrinsically gapless SPT phase. Indeed, this string order cannot have long-range order in a gapped phase\footnote{See Appendix \ref{appgappedstring} for a direct proof.}.
%To see that the system is indeed non-trivial with just $\bZ_4$: the string which has long-range order, $\cdots P_{n-2} P_{n-1} S^z_n$, is charged under $R_x$. Hence, as long as we preserve $R_x$, this discrete charge is well-defined\footnote{Recall that we automatically have $P$ as a symmetry, either by appealing to the fact that it is an unbreakable fermionic symmetry, or by the fact that $R_x^2 = P$---hence one can realize the same physics in a purely bosonic system with $\bZ_4$ symmetry, see Section \ref{secfieldtheory} below.}.
%Our phase is therefore an intrinsically gapless topological phase, as such
%This behavior is not possible in a gapped system, since it would contradict the classification\footnote{See Appendix \ref{appgappedstring} for a direct proof.}.
% Indeed, more generally, if one had a gapped SPT phase with a self-charged string order, i.e., where the end of the string transforms under the  symmetry generator that builds the string, one would obtain a non-trivial SPT phase for the cylic group generated by that symmetry, in contradiction with $H^2(\mathbb Z_n,U(1))=0$.

To avoid confusion, let us note that we call a system topological (with respect to a symmetry group $G$) if its topological phenomena can only be destroyed by (i) explicitly breaking the protecting symmetry or (ii) changing the bulk universality class (either by (iia) tuning off criticality, or (iib) tuning through a multicritical point). In this sense, the above model is a (gapless) topological phase with respect to $\mathbb Z_4$. However, if one also wants the gaplessness itself to be a robust property, we might call it a \textit{stable} gapless (topological) phase. E.g., the above topological Luttinger liquid is stable if we preserve translation symmetry in addition to $\mathbb Z_4$; this can be inferred from measuring its Luttinger liquid parameter to be smaller than $1/2$, as discussed in Appendix~\ref{appisinghub}.
%the $U(1)$ particle number (see Appendix \ref{appnearbyphasediagram}). In fact, it is sufficient to preserve its $\mathbb Z_6 \cong \mathbb Z_2^F \times \mathbb Z_3 \subset U(1)$ subgroup (i.e., preserve fermions modulo six). In other words, the doped Ising chain is a stable gapless SPT phase for the symmetry group $\mathbb Z_4 \times \mathbb Z_3$, whilst there are still no gapped SPT phases for this group.

%To discuss these phenomena, we will use ``topological phase" or ``SPT" in reference to gapless systems to mean a system with edge modes which cannot be removed except by either explicitly breaking the protecting symmetries, driving the system to a multicritical point, or tuning the system off criticality.
%We will use the phrase ``stable gapless phase" to refer to any gapless system, topological or not, for which one cannot open a gap without first passing through a multicritical point or explicitly breaking the protecting symmetries. Finally, an intrinsically gapless topological phase is a gapless SPT whose edge modes cannot be cancelled by symmetric coupling to a gapped SPT edge mode. For our 1d examples, this is always seen by the exotic charges of the string order parameters.

%Since it is not a two-point function as in the Ising phase,
The topology of this gapless phase appears in the form of edge modes. Unlike in the Ising phase,
long-range order in this string operator does not lead to a bulk symmetry breaking degeneracy. Instead, it leads to degeneracy on an interval with open boundaries. To see this, note that the string will still have long-range order even as its endpoints approach the boundaries. Applying the global symmetry transformation $P$, the correlator then becomes a correlation function for a pair of \emph{local} charged operators localized at each boundary. By locality, this means that each boundary has a spontaneous expectation value for $\langle S^z_n\rangle \neq 0$. Summarizing this schematically: $0 \neq \langle S^z_1 P_2 \cdots P_{N-1}S^z_N \rangle = \pm \langle P_1 S^z_1 P_N S^z_N\rangle $ implies $\langle P_1 S^z_1\rangle \neq 0$ by clustering.
This leads to an exponentially split ground state degeneracy, with correlation length set by the spin gap. The energy splitting of the bulk spectrum is much larger at $\sim 1/L$, so this degeneracy can be detected in the finite size spectrum as sketched in Fig. \ref{fig:dopedIsing}(c).

The special properties of this phase can be described in terms of an \textit{emergent anomaly}. Indeed, since fermions are gapped, the parity subgroup $\mathbb Z_2 \subset \mathbb Z_4$ only acts on gapped degrees of freedom, such that $R_x$ acts as a $\bZ_4/\bZ_2 \cong \bZ_2$ symmetry on the low-energy theory. In fact, we will see that its action is incompatible with an on-site microscopic $\bZ_2$ symmetry, which is the essence of the anomaly. In our model $R_x$ is on-site, but the loophole is that $R_x^2 = (-1)^F$, so the anomaly is only an emergent property of the low energy degrees of freedom.
% See Ref.~\cite{Metlitski_2018} for a recent discussion of emergent anomalies in other settings.

The anomaly is illustrated by $R_x$ string correlators of the form $\langle\cO_m(\prod_{m < k < n} R_x ) \cO_n\rangle$, where $\cO$ is a local operator. This correlation function either tends to zero either algebraically or exponentially quickly. As we will see, it turns that out we are in the algebraic case if and only if $\cO$ has odd fermion parity. Since such operators have a non-trivial charge under $R_x^2$, from the point of view of the gapless degrees of freedom, these $R_x$ strings, associated with an effective $\bZ_2$ symmetry, have fractional charge. This fractional charge is a hallmark of the anomaly, as we describe in more detail in Appendix \ref{appbiganomstring}. We will derive the anomaly from the low energy field theory in Section \ref{secfieldtheory} and argue that emergent anomalies are always associated with edge modes in Section \ref{secgenframe}.

\section{Effective Field Theory}\label{secfieldtheory}

In this section we present a field theory of the intrinsically gapless SPT phase above. To focus on the essential features, we present a stripped-down version, where we start from free spinful fermions ($U=J_z=h_x=0$ above) and consider a single perturbation that drives us into one of two topologically-distinct Luttinger liquids, with one of the two having protected edge modes and an emergent anomaly. The free spinful fermion thus plays the role of a phase transition where the fermion becomes gapless and the emergent anomaly jumps.
%The emergent anomaly associated with this phase is straightforward to recognize, and there is also a nearby trivial gapless phase to contrast it with. \textbf{TODO}: where does the free fermion point sit in the model from the previous section? \rv{Answer: $U=J_z=h_x=0$}

We represent the fermion in abelian bosonization as a pair of $2\pi$-periodic compact boson fields $(\varphi_\uparrow,\theta_\uparrow)$, $(\varphi_\downarrow,\theta_\downarrow)$ satisfying $[\partial_x \varphi_s(y),\theta_{s'}(x)] = 2\pi i \delta_{ss'} \delta(x-y)$, in which the fermions may be expressed as $\psi^\dagger_{s,\pm} = U_s e^{\pm i\varphi_s/2 + i\theta_s}$. Here $\pm$ denote the left and right-movers and $U_{1,2}$ the Klein factors, necessary to make these two operators anticommute \cite{von_Delft_1998}. In this theory, $\phi_s \mapsto \phi_s + 2\pi$, $\theta_s \mapsto \theta_s + \pi$ are gauge symmetries for each spin species, see Appendix \ref{appbosonization}. Our $\bZ_4$ symmetry of interest acts as $\psi^\dagger_{s,\pm} \mapsto i \psi^\dagger_{-s,\pm}$ from which we infer the action on the compact boson fields:
\[\label{eqnsymmactions} 
R_x: \begin{cases}
\varphi_s \mapsto \varphi_{-s} \\
\theta_\uparrow \mapsto \theta_\downarrow +\pi/2 \\
\theta_\downarrow \mapsto \theta_\uparrow-\pi/2  \end{cases} \qquad \begin{cases} U_\uparrow \mapsto U_\downarrow \\ U_\downarrow \mapsto -U_\uparrow \end{cases}\]
i.e. rotation about the $x$ axis exchanges the opposite spin fermions and they acquire a phase so that this satisfies $R_x^2 = (-1)^F$.

The operator which tunes us into the two topologically-distinct Luttinger liquids is $\cO_{zz} = \cos (\varphi_\uparrow - \varphi_\downarrow)$. This pins\footnote{This operator is marginal at the free-fermion point, but it can be made relevant by tuning exactly marginal parameters. See Appendix \ref{appisinghub} or \ref{appnearbyphasediagram} for more details.} the spin field $\Phi_1 = \varphi_\uparrow - \varphi_\downarrow$, and hence all states of odd fermion parity are gapped. The remaining low energy degrees of freedom can be described as a Luttinger liquid of spinless Cooper pairs
\begin{equation}
\psi_{\uparrow,+}^\dagger \psi_{\downarrow,-}^\dagger \sim \exp\left(i(\theta_\uparrow + \theta_\downarrow - \varphi_\uparrow/2 + \varphi_\downarrow/2)\right). \label{eq:cooper}
\end{equation}
We can express this in terms of the conjugate compact boson fields\footnote{We are grateful to Max Metlitski for a discussion about these variables.} $\Phi_2 = \varphi_\downarrow$ and $\Theta_2 = \theta_\uparrow + \theta_\downarrow - \varphi_\uparrow/2 + \varphi_\downarrow/2$ . See Appendix \ref{appothersymmetryclasses} for more details.

To determine the symmetry action on these fields,
%$\Phi_2$ and $\Theta_2$, we first note that $\Theta_2$ is not gauge invariant, so the real low energy bosonic fields are $\Phi_2$ and $\Theta_2 = \Theta_2 + \langle \Phi_1 \rangle/2$. Similarly, by
we use \eqref{eqnsymmactions} and replace $\Phi_1$ by its vev wherever it appears.
%vacuum expectation value (vev) $\langle \Phi_1 \rangle$, we find that
We see that $R_x$ acts as a $\bZ_2$ symmetry in the effective low energy theory:
\[\label{eqneffsymm}
R_x:\begin{cases} \Phi_2 \mapsto \Phi_2 + \langle \Phi_1 \rangle \\
\Theta_2 \mapsto \Theta_2 + \langle \Phi_1 \rangle,
\end{cases}
\]
where $\langle \Phi_1 \rangle = 0$ or $\pi$ depending on the sign of the $\cos \Phi_1$ perturbation. If we take the sign to be positive, then $\langle \Phi_1 \rangle = \pi$ and in this case the action of $R_x$ matches the anomalous action at the boundary of the CZX/Levin-Gu $\bZ_2$-SPT phase (compare to Eq. (53) of \cite{Levin_2012} and also \cite{Chen_2011}). This field theory thus describes an intrinsically gapless SPT phase which is equivalent to the one identified in Section \ref{seclattice}. Indeed, it describes the same CFT, and $R_x$ acts as the unique anomalous $\bZ_2$ symmetry. If we perturb with the negative sign of $\cO_{zz}$ on the other hand, we find a trivial Luttinger liquid phase with $\langle \Phi_1 \rangle = 0$ and trivial $R_x$ action.
In the lattice model in Section \ref{seclattice}, the trivial phase is obtained by driving a different parameter and we cannot capture this alternative transition in this field theory. We have also confirmed the above prediction in a lattice model that closely realizes this field theory; see Appendix~\ref{appisinghub}.
%by a more involved competition between $O_{zz}$ and $O_x = \cos((\varphi_\uparrow+\varphi_\downarrow)/2)\cos(\theta_\uparrow-\theta_\downarrow)$, but this only affects the nature of the multicritical point.

The string order for fermion parity in this theory may be derived from the above two-component Luttinger liquid as follows. By the canonical commutation relations, we see the generator of fermion parity $\theta_{1,2} \mapsto \theta_{1,2} + \pi$ is given by $\exp(i \int dx( \partial_x \varphi_\uparrow/2 + \partial_x \varphi_\downarrow/2))$. To obtain the string operator, we first truncate the integral so it goes from $-\infty$ to $x$, which reduces it to a boundary term. In the new variables it becomes $\exp(i \Phi_1(x)/2 + i \Phi_2(x))$. Note that in either $c = 1$ phase, where $\Phi_1$ is gapped and has a v.e.v., this operator is mutually local with the low energy operators. Indeed fermion parity is a gapped symmetry. However, because of the $e^{i\Phi_2(x)}$, its correlation function has algebraic decay, so we must take the endpoint operator $\cO(x) = e^{-i\Phi_2(x)}$ to cancel this factor and obtain a string with long-range order. This end point operator is charged under $R_x$ in the topological phase (cf. \eqref{eqneffsymm}), just as we observed in Section \ref{seclattice}.

We can see the edge modes by studying a spatial interface from the topological to trivial Luttinger liquid where we tune the coefficient of the $\cO_{zz} = \cos \Phi_1$ perturbation from a positive to a negative value, adapting an argument from Ref.~\cite{Keselman15}. There is an edge mode associated with the path $\Phi_1$ takes from 0 to $\pi$ across the interface.
%Whatever the form of the interface, there will be two generate paths minimizing the potential, which are exchanged by
Any continuous path that minimizes the energy across this interface comes with a degenerate partner by exchanging 
$R_x:\Phi_1 \to - \Phi_1$. One way to see this is to observe that because $\Phi_1$ is pinned to its vev far from the interface, $\langle S^z \rangle = \frac{1}{2} \left\langle \int_{-\infty}^\infty \frac{d\Phi_1}{2\pi} \right\rangle = \pm \frac{1}{4}$, from which we see that the edge mode is a spin-$1/2$ qubit. Together with the gapless charges in the bulk, this implies that at the boundary, fermions become gapless, as we anticipated in Section \ref{seclattice}.
%toggled by occupying or vacating a spin-$\frac{1}{2}$ fermion state at the edge, as we anticipated in Section \ref{seclattice}.

This field theory has  appeared in Refs.~\cite{Keselman15,Kainaris15,Kainaris17} although  the protecting symmetry group there was $U(1) \rtimes T$ instead of our $\mathbb Z_4$. In the former case, the gapped $\mathbb Z_2$ subgroup of $U(1)$ has a string order whose charge is incompatible with any $U(1)$-symmetric gapped SPT. However, if one explicitly breaks the $U(1)$ symmetry down to its $\mathbb Z_2$ subgroup, one obtains a gapped $\mathbb Z_4^T$ SPT phase. In this sense, the $\mathbb Z_4$ example in this work gives a conceptually cleaner instance of an intrinsically gapless SPT phase, as there is not even any subgroup that protects a gapped SPT phase. See also  Table~\ref{table:anom} and Appendix \ref{appothersymmetryclasses}. To the best of our knowledge, the emergent anomaly viewpoint which is central to our discussion of these models, has not appeared  in  earlier works.

\begin{table*}
 \begin{tabular}{||c c c c c c c||} 
 \hline
 $d$ & $G$ & $G_{\rm low}$ & comment & $\omega$ & $\alpha$ & SPT$_d$ subgroup \\ [0.5ex] 
 \hline\hline
 1 & $\bZ_4$ & $\bZ_2$ & Levin-Gu/CZX anomaly & $\frac{1}{4} A_{\rm low} dA_{\rm low}$ & $\frac{1}{2}A_{\rm gap} A_{\rm low}$ & none \\
 \hline
 1 & $U(1) \times \bZ_2$ & $U(1) \times \bZ_2$ & bosonic QSH anomaly & $\frac{1}{4\pi} A_{\rm low}^{\bZ_2} dA_{\rm low}^{U(1)} $ & $\frac{1}{2} A_{\rm gap} A_{\rm low}^{\bZ_2}$ & $\bZ_2 \times \bZ_2$  \\
 \hline
 1 & $Pin^-(2)$ & $U(1) \rtimes \bZ_2^T$ & bosonic TI anomaly & $\frac{1}{4\pi} w_1 dA_{\rm low}^{U(1)}$ & $\frac{1}{2} w_1 A_{\rm gap}$ & $\bZ_4^T$  \\
 \hline
 2 & $Spin^c(3)$ & $SO(3) \times U(1)$ & Neel-VBS DQCP & $\frac{1}{2} w_4(A_{\rm low})$ & $\frac{1}{4\pi} A_{\rm gap}dA_{\rm low}^{U(1)} + \cdots$ & $U(1)$ or $SU(2)$ \\[1ex] 
  \hline
%  $d$ & $SU(2) \times \bZ^d$ & $SO(3) \times \bZ^d$ & LSMOH theorem & $\frac{1}{2} w_2(A_{\rm low}^{SO(3)}) \tau_1 \cdots \tau_d$ & $\frac{1}{2} A_{\rm gap}\tau_1 \cdots \tau_d$ & $\bZ_2^F \times \bZ^d$ \\[1ex] 
%  \hline
\end{tabular}
\caption{\textbf{Emergent anomalies in intrinsically gapless SPT phases (protected by $G$ in $d$ spatial dimensions).} These examples can occur in fermionic systems where all odd-parity states are gapped (or alternatively in a bosonic system with $G_{\rm gap} = \mathbb Z_2$). The details of these calculations can be found in Appendix \ref{subappexamples}. The case $G = \bZ_4$ was discussed in Sections \ref{seclattice} and \ref{secfieldtheory} above. Meanwhile, $G = U(1) \times \bZ_2$ captures the phenomena of the gapless Haldane phase in Ref.~\cite{Kestner11} or the charge-conserving topological superconductors in Refs.~\cite{Cheng11,Kraus13,Iemini15,Lang15,Guther17,Chen18} and $G = Pin^-(2)$, generated by $U(1)$ charge and $T^2 = (-1)^F$, that of Refs.~\cite{Keselman15,Kainaris15,Kainaris17}. We describe the field theory for these examples in Appendix \ref{appothersymmetryclasses}. Unlike $G = \bZ_4$, the robustness of the edge modes in these models can be understood by considering a subgroup of $G$, namely $\bZ_2 \times \bZ_2$ and $\bZ_4^T$ respectively. In the latter case, the 1st Stiefel-Whitney class $w_1$ (of the tangent bundle of $X$) appears in both $\omega$ and $\alpha$, which plays the role of a time-reversal gauge field \cite{kapustin2014symmetry,kapustin2014bosonic}. The anomaly of the %Neel-VBS 
deconfined quantum critical point (DQCP),  \cite{Senthil04,Wang_2017,Komargodski_2019,Metlitski_2018} can also be cured by embedding the system into a fermionic Hilbert space with rotation and charge conservation (i.e. $Spin^c(3)$ or even $Spin(5)$ symmetry) realized as on-site symmetries \cite{Grover08,Ippoliti18,QSHDQCP,MikeZilla}. In this case, there are several possible solutions to the anomaly vanishing equation, indicated by $\cdots$ in $\alpha$, which essentially contain $Spin^c(3)$ Chern-Simons terms. In the anomaly $w_4$ is obtained by restriction of the 4th Stiefel-Whitney class by the block-diagonal embedding $SO(3) \times U(1) \subset SO(5)$.}
\label{table:anom}
\end{table*}

\section{General Framework}\label{secgenframe}

To recap, the key ingredients of an intrinsically gapless phase of the sort explored above is (i) a system with an on-site symmetry (e.g., $R_x$), (ii) a gapless phase where part of the symmetry (e.g., $(-1)^F$) acts only on gapped degrees of freedom and (iii) the action of the remaining symmetry on the gapless degrees of freedom is anomalous, i.e. an emergent anomaly.
%One can detect the emergent anomaly by measuring the charges of string order parameters. Although this procedure may seem analogous to an SPT phase,  the charges obtained are incompatible with any gapped phase. Furthermore, the charged string operators imply the existence of edge modes at a symmetric boundary.
In 1+1D, one can detect the emergent anomaly by measuring the charges of string order parameters---note that unbroken symmetries that act purely on gapped degrees of freedom always admit a string order parameter (see Appendix~\ref{appbiganomstring}). The  non-trivial charges of the latter also imply edge modes, which seems analogous to an SPT phase, but the charges themselves are incompatible with any gapped phase.

Let us now describe a general picture of this phenomenon, which applies to a general symmetry group and any dimension. Let $G$ be the microscopic on-site symmetry, $G_{\rm gap}$ be the (normal) subgroup of $G$ which acts trivially on the gapless degrees of freedom, $G_{\rm low} = G/G_{\rm gap}$ be the effective symmetry of the low energy theory, with $\pi:G \to G_{\rm low}$ being the quotient map.

For any system with such a gapped sector but a gapless low energy theory, if we compute the partition function $Z(X,A)$ of this system on a spacetime $X$ coupled to background $G$ gauge field $A$\footnote{Our discrete gauge fields can be interpreted as simplicial cochains, where $d$ is the simplicial differential, and products are cup products. Thus they generalize the familiar $U(1)$ gauge potentials with the extra provisio that for e.g. $G=\bZ_N$ they are valued in integers mod $N$, so that the bosonic Chern-Simons term $\frac{1}{2\pi}A\wedge dA$ becomes $\frac{2\pi}{N^2} A\cup dA$. See Appendix \ref{appgaugefield} for more details.}, we will find it has the form
\[\label{eqnpartfun}Z(X,A) = Z_{\rm low}(X,A_{\rm low})e^{2\pi i \int_X \alpha(A)} + \cdots,\]
where $Z_{\rm low}(X,A_{\rm low})$ is the partition function of the gapless degrees of freedom coupled to background $G_{\rm low}$ gauge field derived from $A$ by $A_{\rm low} = \pi(A)$, $\alpha(A)$ is a topological term obtained after integrating out the gapped degrees of freedom, and $\cdots$ are terms exponentially small in the gap that we now discard. In the case that there is no emergent anomaly, $Z_{\rm low}(X,A_{\rm low})$ and $\alpha(A)$ are both gauge invariant, and $\alpha(A)$ describes a $G$-SPT phase in the gapped sector.

When there is an emergent anomaly, on the other hand, $Z_{\rm low}(X,A_{\rm low})$ and $\alpha(A)$ are not separately gauge invariant, and instead transform in such a way that only their combination $Z(X,A)$ is gauge invariant. We cannot interpret $\alpha$ as an SPT class in this case. Instead, invoking the bulk-boundary correspondence, we can express the emergent anomaly in terms of a topological term $\omega(A_{\rm low})$ for a $G_{\rm low}$ SPT in one higher dimension \cite{Chen_2013,kapustin2014anomalies}. This means that for $\partial X = 0$, $Z_{\rm low}(X,A_{\rm low})\exp\left(2\pi i \int_{X \times \mathbb{R}_{\ge 0}} \omega(A_{\rm low})\right)$ is gauge invariant.
By standard arguments, gauge invariance of \eqref{eqnpartfun} on closed spacetime manifolds is then equivalent to the anomaly vanishing equation $d\alpha = \omega$ (see Appendix \ref{appgaugefield}).

We see that for this equation to be solvable, $\alpha(A)$ has to depend on the $G_{\rm gap}$ part of $A$, since otherwise $\omega$ would describe a trivial $G_{\rm low}$ SPT. This means that when we perform a $G_{\rm gap}$ gauge transformation, $\alpha(A)$ will shift by an exact but nonzero form, meaning $\alpha(A+dg) = \alpha(A) + d\lambda(A,g)$ for some $\lambda$, while $Z(X,A_{\rm low})$ will remain unchanged. When we study the partition function on a spacetime $X$ with boundary, this will lead to a boundary term $e^{2\pi i \int_{\partial X}\lambda(A,g)}$, indicating that there must be an extra boundary contribution which makes the combination \eqref{eqnpartfun} gauge invariant again. This extra boundary contribution must come from some kind of edge mode.

Let us illustrate the above for our $G = \bZ_4$ example, where $G_{\rm gap} = \bZ_2$ and $G_{\rm low} = \bZ_2$. The $\bZ_2$ anomaly has the Chern-Simons form $\omega(A_{\rm low}) = \frac{1}{4} A_{\rm low} dA_{\rm low}$ \cite{kapustin2014anomalies}. We can write the $\mathbb Z_4$-valued $A$ as a combination of $\bZ_2$ gauge fields $A_{\rm low}$ and $A_{\rm gap}$ according to $A = 2A_{\rm gap} + A_{\rm low}$, with the extension to $\bZ_4$ encoded into the equation $dA = 0$ mod 4 $\Leftrightarrow$ $2dA_{\rm gap} = dA_{\rm low}$ mod 4. This equation says that a $2\pi$-flux of $A_{\rm low}$ equals a $\pi$-flux of $A_{\rm gap}$, which captures the relation $R_x^2 = (-1)^F$. We find the only topological term which satisfies the anomaly vanishing equation is $\alpha(A) = \frac{1}{2} A_{\rm gap} A_{\rm low}$. Under a gauge transformation $A_{\rm gap} \mapsto A_{\rm gap} + dg$, there is a boundary term $\frac{1}{2} g \; A_{\rm low}$, as claimed. This implies the existence of an edge mode to restore gauge invariance.

This argument can be related to the 1d argument for edge modes based on string operators. Indeed, one may observe that $\alpha(A) = \frac{1}{2} A_{\rm gap}  A_{\rm low}$ in the example above means that the fermion parity string has an endpoint operator charged under $R_x$, as we have observed in Sections \ref{seclattice} and \ref{secfieldtheory}. This is because the fermion parity string corresponds to the ground state in the parity twisted sector, which has $\int_{\rm space} A_{\rm gap} = 1$ mod 2. The topological term $\alpha$ thus contributes 1 to the total $A_{\rm low}$ charge. Likewise we find that every state in the $R_x$-twisted sector of the low energy theory has odd fermion parity, as we claimed in Section \ref{seclattice}. See Appendix \ref{appemerganom} for more details and examples, summarized in Table~\ref{table:anom}.

%\av{Next sentence needs to be fixed. i also moved the rest to the footnote, since its already a lot to digest for a first reading. }
%This approach let's us state a kind of classification statement, which is

This approach also tells us that \textit{up to stacking with a $G$-SPT}, the partition function of the gapless SPT is classified by the low energy theory and its emergent anomaly (we do not know if two theories with the same $Z(X,A)$ can be deformed into one another, although it is a standard assumption in gapped classifications \cite{kapustin2014symmetry,KTTW}). Indeed, in general there may be different $\alpha$ that solve the anomaly vanishing equation. If $\alpha$ and $\alpha'$ are two solutions, then $d(\alpha - \alpha') = 0$ so $\alpha - \alpha'$ describes a $G$-SPT which can be interpreted as relating the two topological phases by stacking\footnote{Therefore, if there is one solution to the anomaly vanishing equation, then there are as many as there are $G$-SPTs in the same dimension. It may be that not all of these describe distinct phases however, since often the gapless system can ``absorb" an SPT involving only the gapless symmetries \cite{verresen_gapless_2019}. For instance, if $G_{\rm gap} = 1$, so there is no anomaly and all symmetries are gapless, it can happen that even if $\alpha \neq 0$, there are no protected edge modes. This illustrates the importance of using the $G_{\rm gap}$ gauge transformations in the argument above.}. 

Another viewpoint on our work  emerges by thinking about `failed' 2+1D bosonic SPT phases, i.e. those that can be trivialized by embedding them into a bigger Hilbert space---a natural instance being the addition of fermions \cite{Lu12,Lu14,Wang14,Gu14,Wang_2016}---meaning that the edge theory is no longer absolutely protected.
The present work shows that there is still a well-defined emergent anomaly at the edge, and moreover it can therefore be realized in its own dimension as an intrinsically gapless topological phase.

Thus far, we have studied various examples of intrinsically gapless topological phases, but how do we systematically construct them? This question is answered in Appendix \ref{subappconstruction} where we discuss the construction of intrinsically gapless SPT on the lattice, beginning  from an anomalous theory (i.e., a gapless edge of an SPT lattice model) and a solution to the anomaly vanishing equation, i.e. a prescription to augment the degrees of freedom to `cure' the anomaly. Moreover, adapting the results of \cite{Tachikawa_2020,Wang_2018,thorngren2020tqft} (who studied the problem of constructing gapped symmetric boundaries of SPTs and encountered the same equation $d\alpha = \omega$), for any theory with a bosonic global $G_{\rm low}$ anomaly, there is some $G_{\rm gap}$ which realizes it as an intrinsically gapless phase where the anomaly is completely emergent.

%one can prove a sort of converse of this classification, to go from an anomalous theory (i.e. a gapless edge of an SPT lattice model) and a solution to the anomaly vanishing equation to an intrinsically gapless SPT on the lattice.

\section{Outlook}

% Several gapless topological phases in the literature can be interpreted as intrinsically gapless SPT phases. A bosonic example is given by the gaples Haldane phase protected by $U(1) \times \mathbb Z_2$ \cite{Kestner11}, whereas a multitude of fermionic examples arose in the quest of finding Majorana-like edge states in number-conserving models, leading to gapless SPTs protected by $U(1)\times \mathbb Z_2$ \cite{Cheng11,Kraus13,Iemini15,Lang15,Guther17,Chen18} or $U(1) \rtimes T$ \cite{Keselman15,Kainaris15,Kainaris17}. In all these cases, the gapped $\mathbb Z_2$ subgroup of $U(1)$ has a string order whose charge is incompatible with any $U(1)$-symmetric gapped SPT. However, if one explicitly breaks the $U(1)$ symmetry down to its $\mathbb Z_2$ subgroup, one obtains a gapped SPT phase. In this sense, the $\mathbb Z_4$ example in this work gives a cleaner instance of an intrinsically gapless SPT phase. We describe the field theory for these other symmetry classes in Appendix \ref{appothersymmetryclasses}.

%Anomalies are one of the cornerstones of recent developments in high-energy and condensed matter physics.
Anomalies are a key nonperturbative phenomenon in high-energy and condensed matter physics \cite{adler,belljackiw,tHooft,Alvarez86,Witten_2016}. However, they can be difficult to realize, either living on the boundary of a higher-dimensional system or requiring a non-internal symmetry action. In this work, we have shown how they can arise from \emph{on-site} symmetries in the \emph{same} dimension and how---in one dimension---they can be diagnosed using unusual string order. This paves the way for experimental realization of anomalies. In fact, most of the machinery is already in place. Doped spin-$1/2$ Heisenberg chains \cite{Hilker17} have been realized and their string order has been measured---although the latter decayed algebraically due to spin-rotation symmetry. If one can engineer an Ising anisotropy, then the lattice model in Section~\ref{seclattice} is realized, which should endow the string order parameter with long-range order. As discussed, the latter implies edge modes with $S^z = \pm \frac{1}{4}$, whose fractional value should be measurable by considering statistical ensembles.

The framework introduced in this work  provides guidance to constructing interesting new models, not just in 1+1D but also in higher dimensions. 
%Other exciting directions for future exploration are higher-dimensional examples. As discussed above, is applicable to any dimension. 
A paradigmatic example of an anomaly in 2+1D is at a deconfined quantum critical point, which has been proposed to describe a transition between a Heisenberg antiferromagnet and a valence bond solid \cite{Senthil04}. Here, SO(3) spin rotation and an effective U(1) symmetry arise, despite the anomaly, by utilizing spatial symmetries to implement (a discrete subgroup of) the latter symmetry.  Alternatively, the U(1) can be identified with an internal symmetry related to charge conservation. In this case the anomaly of the deconfined critical point can be lifted by embedding it into a fermionic Hilbert space \cite{Grover08,Ippoliti18,QSHDQCP,MikeZilla}; see Table~\ref{table:anom}. Our theory predicts that with on-site SO(3) and U(1) symmetries, implemented  at the expense of including gapped fermions, an exotic 1+1D edge theory will appear at the gapless deconfined critical point. This direction demands further study.

Finally, although we have highlighted one mechanism for intrinsically gapless topological phases, we do not know if it is the only mechanism. In particular, because the emergent anomaly relies on there being a gapped symmetry, one can ask if there are also intrinsically gapless SPTs for which the whole symmetry group is gapless, or in models with no gapped sector at all.

\begin{acknowledgments}
We thank Dave Aasen, Immanuel Bloch, Nick G. Jones, Max Metlitski, Dan Parker, Nat Tantivasadakarn and Yifan Wang for useful discussions, and especially Dan, Max, Nick, and Yifan for a careful reading of the manuscript. RV is indebted to Pablo Sala for a very fruitful discussion that gave the inspiration for this work when we realized that the $t-J_z$-chain discussed in Ref.~\cite{Rakovszky20} might have an unusual symmetry protection. The MPS-based DMRG simulations were performed using the Tensor Network Python (TeNPy) package developed by Hauschild and Pollmann \cite{Hauschild18}. This work was supported by the Harvard Quantum Initiative Postdoctoral Fellowship in Science and Engineering (RV) and a grant from the Simons Foundation (\#376207) (AV, RV).
\end{acknowledgments}

%bosonic: $U(1) \times \mathbb Z_2$ \cite{Kestner11}
%fermionic: $U(1) \times \mathbb Z_2$ \cite{Cheng11,Kraus13,Iemini15,Lang15,Guther17,Chen18}
%fermionic: $U(1) \rtimes \mathbb Z_2^T$ \cite{Keselman15,Kainaris15,Kainaris17}

%\textbf{TODO}\av{make some comment about simulating anomalous phases in the same dimension with internal symmetries? Experiments - Bloch realization of Ising Hubard (already has the tools in this paper and can measure the string order parameters accoring tot his paper, edge states??)}

%\av{Could the effect of disorder be interesting - may remove the need for extra symmetries and symmetries are on-site. }
\bibliography{refs.bib}

\onecolumngrid
\appendix

\section{More numerical results \label{appisinghub}}

\subsection{The Ising-Hubbard chain}

\subsubsection{Confirming criticality and topology}
\begin{figure}[h]
    \centering
    \includegraphics{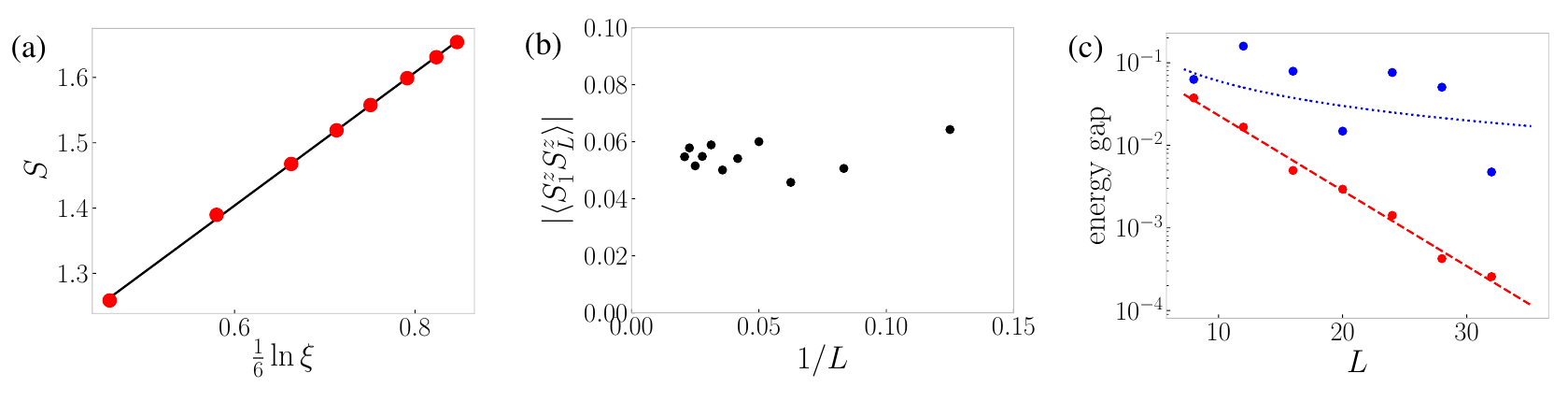}
    \caption{The Ising-Hubbard chain for $\mu=0$, $h_x=0.1$, $J_z=1=t$ and $U=20$. (a) The entanglement scaling indicates a conformal field theory with central charge $c_{\rm fit} \approx 1.02$. (b) The boundaries are spontaneously magnetized, confirmed by the long-range order in the boundary-boundary correlator. (c) The finite-size splitting is exponentially small in system size (red). The blue dots are the gap to the next excited state; the blue dotted line is $\sim 1/L$ as a guide to the eye. \label{fig:appisinghub}}
\end{figure}

To confirm that we indeed have a Luttinger liquid, we extract the central charge $c=1$ using entanglement scaling \cite{Calabrese04,Pollmann09}, plotted in Fig.~\ref{fig:appisinghub}(a). We have already numerically confirmed that the system does not spontaneously break the Ising symmetry; see the algebraically decaying spin correlations in Fig.~\ref{fig:dopedIsing}(b). However, with open boundaries, the edge-to-edge spin correlation function has long-range order (Fig.~\ref{fig:appisinghub}(b)), indicating that the boundaries spontaneously break $R_x$ symmetry. The resulting degeneracy has an exponentially small finite-size splitting for finite systems, shown by the red dots in Fig.~\ref{fig:appisinghub}(c). This is significantly smaller than the bulk $\sim 1/L$ energy scale. The scaling of the latter (blue dots in Fig.~\ref{fig:appisinghub}(c)) is significantly affected by Friedel oscillations.

\subsubsection{Lattice-continuum correspondence and Luttinger liquid parameter}

Here we build a correspondence between the lattice model of Section~\ref{seclattice} and the field theory discussed in Section~\ref{secfieldtheory}. This will also allow us to numerically extract the Luttinger liquid parameter of the topological phase, which in turn carries information about the stability of the phase.

Let us first consider the Ising order parameter $S^z_j$. We have already seen in Fig.~\ref{fig:dopedIsing}(b) that this decays algebraically. To determine which CFT operator it generates, we write (suppressing momentum-dependent prefactors):
\begin{equation}
    S^z_j \sim \partial(\varphi_\uparrow-\varphi_\downarrow) + \left( e^{i\varphi_\uparrow} - e^{i\varphi_\downarrow}\right) + \cdots = \partial \Phi_1 + e^{i \Phi_2} \left( e^{i \Phi_1} - 1\right) + \cdots.
\end{equation}
Hence, in the trivial phase, where $\Phi_1=0$, we obtain that $S^z_j \sim 0$. Indeed, this must happen at all orders, since $S^z_j$ is odd under $R_x$, whereas in Section~\ref{secfieldtheory} we have seen that the low-energy fields of the trivial Luttinger liquid are all even under $R_x$. We have numerically confirmed that in the trivial phase, $\langle S^z_i S^z_j \rangle$ decays exponentially fast.

\begin{figure}
    \centering
    \includegraphics[scale=0.35]{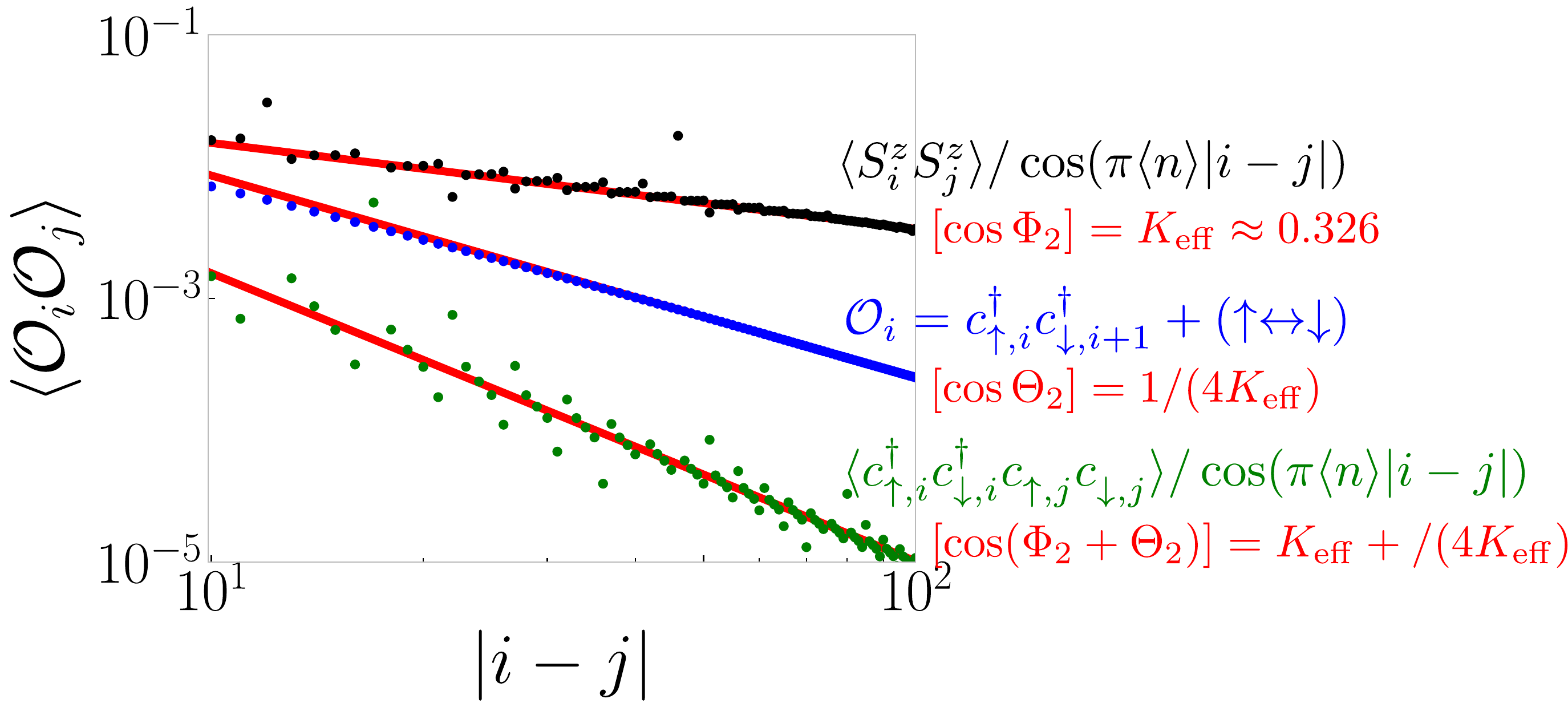}
    \caption{Lattice operators in the Ising-Hubbard chain and the corresponding low-energy CFT operators that they generate in the topological Luttinger liquid. We use one of these to extract the Luttinger liquid parameter $K_{\rm eff}$; the other two then give strong consistency checks. We have multiplied some of these correlation functions by an overall constant prefactor to shift the curves up for easier presentation.}
    \label{fig:correspondence}
\end{figure}

In the topological phase, we have $\Phi_1=\pi$, such that $S^z \sim \cos \Phi_2$. We can thus read off the effective Luttinger liquid parameter $K_{\rm eff}$ from $\langle S^z_i S^z_j \rangle \sim 1/|i-j|^{2K_{\rm eff}}$. Since $\Phi_2 = \varphi_\downarrow$, this will carry momentum $\langle n \rangle \pi$, where $\langle n \rangle$ is the particle filling. Fixing $t=J_z=1$, $\mu=\frac{1}{2}$ and $U=5$ (and $h_x=0$)---i.e., the same parameters as in Fig.~\ref{fig:dopedIsing}(b)---we measure $\langle n \rangle \approx 0.7935$. Dividing out this oscillatory phase factor, we obtain a straight line in the log-log plot of Fig.~\ref{fig:correspondence}, from which we extract $K_{\rm eff} \approx 0.326$.

We are thus in a strongly-repulsive regime. In particular, all symmetry-allowed $U(1)$-breaking terms (i.e., $\cos(2n \Theta_2)$ and $\sin(2n\Theta_2)$ since $R_x$ shifts $\Theta$ by $\pi$; see Eq.~\eqref{eqneffsymm}) are RG-irrelevant. The relevant symmetry-allowed operators are $\cos(2\Phi_2)$ and $\sin(2\Phi_2)$, with dimensions $4K_{\rm eff}$. Since these carry incommensurate momentum, the gapless phase is stable if we preserve translation symmetry---in addition to the $R_x$ symmetry. For a discussion of the nearby phases generated by the above operators, see Appendix~\ref{appnearbyphasediagram}.

To obtain a lattice operator that generates $\cos(\Theta_2)$, it is natural to consider Eq.~\eqref{eq:cooper}, which we reproduce here for convenience:
\begin{equation}
\psi_{\uparrow,+}^\dagger \psi_{\downarrow,-}^\dagger \sim \exp\left(i(\theta_\uparrow + \theta_\downarrow - \varphi_\uparrow/2 + \varphi_\downarrow/2)\right) = e^{i\Theta_2}.
\end{equation}
One could wonder whether the lattice operator $c^\dagger_{\uparrow,i} c^\dagger_{\downarrow,i}$ generates this field. However, it cannot: this lattice operator is manifestly $R_x$-symmetric, whereas $e^{i\Theta_2}$ is odd under $R_x$ (at least in the topological phase). Indeed, instead one finds that it has contributions of the sort
\begin{equation}
\psi_{\uparrow,+}^\dagger \psi_{\downarrow,+}^\dagger \sim e^{i(\frac{\varphi_\uparrow + \varphi_\downarrow}{2} + \theta_\uparrow + \theta_\downarrow)} \sim e^{i\Phi_1}e^{i(\Theta_2+\Phi_2)}.
\end{equation}
In conclusion, $c^\dagger_{\uparrow,i} c^\dagger_{\downarrow,i} +h.c. \sim \cos (\Theta_2 + \Phi_2)$, which has scaling dimension $K_{\rm eff} + \frac{1}{4K_{\rm eff}}$. This prediction is confirmed in Fig.~\ref{fig:correspondence}.

To generate $\cos \Theta_2$, we must thus consider a lattice operator that is odd under $R_x$. A simple tweak of the above is $c^\dagger_{\uparrow,i} c^\dagger_{\downarrow,i+1} + (\uparrow \leftrightarrow \downarrow)$. We indeed confirm that this has dimension $1/(4K_{\rm eff})$, shown in Fig.~\ref{fig:correspondence}. Note that this operator has no momentum, in contrast to the $\Phi_2$ fields.

\subsection{Coupled Luttinger liquids}\label{subappcoupledluttingers}

Instead of the Ising-Hubbard chain in Eqs.~\eqref{eq:Ising} and \eqref{eq:Hub}, we can also consider two decoupled interacting Luttinger liquids (for each spin species) which are coupled by the Ising term:
\begin{equation}
    H = H_\uparrow + H_\downarrow + J_z \sum_n S^z_j S^z_{j+1} \qquad \textrm{where } H_s = - t \sum_n \left( c^\dagger_{j+1,s} c_{j,s}^{\vphantom \dagger} +  h.c. \right) + V \sum_n n_{j,s} n_{j+1,s},
\end{equation}
with $S^z_j = \frac{1}{2} \left( n_{j,\uparrow} - n_{j,\downarrow} \right)$. This set-up is closest to the field theory discussed in Section~\ref{secfieldtheory}. In particular, for $V=J_z=0$, we have our spinful free-fermion starting point. The Ising term contains $\cos(\varphi_\uparrow-\varphi_\downarrow)$, however, at the free-fermion point this has dimension $K_\uparrow+K_\downarrow=2$, i.e., it is is marginal. Therefore we include the Luttinger interaction which tunes the $R_x$-symmetric $(\partial \varphi_\uparrow)^2 + (\partial \varphi_\downarrow)^2$, such that for repulsive $V>0$, the Luttinger liquid parameter $K_s<1$, making $J_z \cos(\varphi_\uparrow-\varphi_\downarrow)$ relevant. Hence, for such a fixed value of $V>0$, the field theory in Section~\ref{secfieldtheory} predicts that depending on the sign of $J_z$, we get a topological or trivial Luttinger liquid with central charge $c=1$, separated by a $c=2$ phase transition at $J_z=0$.

This prediction is straightforwardly verified. Note that for $J_z=0$, the model is integrable, and using the exact solution we can calculate the resulting Luttinger liquid parameter \cite{LLparameter}. We fix $V=10$, for which $K_\uparrow = K_\downarrow \approx 0.5266$, such that the Ising coupling is relevant with dimension $\approx 1.0532$. As before, we numerically diagnosed criticality by observing $c=1$ from entanglement scaling; the topological and trivial Luttinger liquids were distinguished by measuring long-range order of the topological string-order ($\cdots P_{j-2}P_{j-1}S^z_j$) or the trivial string-order ($\cdots P_{j-2} P_{j-1} P_j$) parameters. We find that the topological and trivial Luttinger liquids  persist over a wide range of parameter space: the phases are observed for $0<|J_z| < 4$ (and slightly beyond), with $J_z>0$ ($J_z<0$) being topolgical (trivial), consistent with the field theory.

\section{Anomalies and String Order Parameters}\label{appbiganomstring}

$G$-Anomalies of 1d systems are in correspondence with 2d $G$-SPTs via the bulk-boundary correspondence \cite{Chen_2013,kapustin2014anomalies}. The latter have been classified by the topological terms which are generated when the SPT is coupled to a background gauge field \cite{dijkgraafwitten}. These topological terms in turn determine the anomalous behavior of string order parameters (a.k.a. symmetry fluxes) in the anomalous 1d theory. In this appendix we describe these anomalous behaviors and how one can use the topological terms to further learn about our gapless phases with emergent anomaly.

\subsection{String Operators in Gapped Phases}\label{appgappedstring}

Before diving into the anomalous case, let us discuss string operators in 1d gapped phases \cite{dennijs89,Perez08,Haegeman12,Pollmann12}. We consider on-site unitary symmetries, meaning
\[U = \prod_x U_x,\]
where the product is over sites $x$ and $U_x$ is a unitary operator acting only on the local Hilbert space of the site and satisfying the group law. If we have a gapped symmetric ground state $|0\rangle$, it is known that if we apply the symmetry to the ground state in a finite region, it is equivalent applying exponentially localized operators near the ends, meaning
\[\label{eqnsymmfrac}\prod_{j \le x \le k} U_x |0\rangle = U_{j,L} U_{k,R} |0\rangle,\]
where $U_{j,L}$ and $U_{k,R}$ are unitary operators with support exponentially localized near $j$ and $k$ respectively. This is known as symmetry fractionalization \cite{Pollmann10,Turner11,Verresen17}.

By conjugation with $U$, we can assume $U_L$ and $U_R$ have some fixed charge $q_L$, $q_R$ under $U$. Let us now show this charge is trivial. We consider the product of two semi-infinite strings (although the same argument can be carried out for finite strings, just with more bookkeeping):
\[(\prod_{x \le n} U_x^\dagger)(\prod_{y \le m} U_y)|0\rangle\]
with $m \gg n$. We have
\[(\prod_{x \le n} U_x^\dagger)(\prod_{y \le m} U_y)|0\rangle = \prod_{n \le y \le m} U_y |0\rangle = U_{n,L} U_{m,R}|0\rangle\]
as well as
\[(\prod_{x \le n} U_x^\dagger)(\prod_{y \le m} U_y)|0\rangle = (\prod_{x \le n} U_x^\dagger) U_{m,R}|0\rangle = U_{n,R}^\dagger U_{m,R}|0\rangle,\]
so $U_L|0\rangle = U_R^\dagger|0\rangle$. We also have
\[(\prod_{x \le n} U_x^\dagger)(\prod_{y \le m} U_y)|0\rangle = (\prod_{y \le m} U_y)(\prod_{x \le n} U_x^\dagger)|0\rangle = (\prod_{y \le m} U_y) U_{n,L} |0\rangle = e^{i q_L} U_{n,L} (\prod_{y \le m} U_y)|0\rangle = e^{i q_L} U_{n,L} U_{m,R}|0\rangle.\]
This proves the claim.

Rearranging \eqref{eqnsymmfrac}, we find
\[\langle 0 |U_{j,L} (\prod_{j \le x \le k} U_x) U_{k,R}|0\rangle = 1,\]
so symmetry fractionalization implies that the string operator $(\prod_{x \le k} U_x) U_{k,R}$ has long-range order. Although $U_R$ cannot be charged under $U$, if there are other global symmetries, then it can be charged, which signals a nontrivial SPT phase. Note that this charge is unique unless the other global symmetries are broken, since otherwise with $U_{R,x} U_{R,x'}'$ two differently charged end operators, we could consider $U_R$, $U_R'$ inserted at some large (relative to the correlation length) but fixed distance $|x - x'|$, and this would be an order parameter for a broken symmetry. As long as the gap remains open and all symmetries are unbroken, these charges give topological invariants for 1d gapped phases.

Another useful fact is that if $U$ and $V$ are commuting symmetries with $U_R$ having charge $q$ under $V$ and $V_R$ having charge $q'$ under $U$, then $q = - q'$ (i.e., there is charge reciprocity). To see this, consider
\[(\prod_{y \le m} V_y) (\prod_{x \ge n} U_x) |0\rangle\]
with $m \gg n$. We have
\[(\prod_{y \le m} V_y) (\prod_{x \ge n} U_x) |0\rangle = (\prod_{y \le m} V_y) U_{R,n}^\dagger |0\rangle = e^{-iq}U_{R,n}^\dagger V_{R,m} |0\rangle\]
On the other hand
\[(\prod_{y \le m} V_y) (\prod_{x \ge n} U_x) |0\rangle =  (\prod_{x \ge n} U_x) (\prod_{y \le m} V_y)|0\rangle = (\prod_{x \ge n} U_x) V_{R,m} |0\rangle = e^{iq'}V_{R,m} U_{R,n}^\dagger |0\rangle.\]
The claim follows.

For abelian discrete symmetries, the charges of these string operators (subject to no self-charges and charge reciprocity) are known to characterize all SPTs. For example with $\bZ_2^U \times \bZ_2^V$, there are two possibilities consistent with charge reciprocity: either both strings have trivial charges or the $U$ string is odd under $V$ and even under $U$ and vice versa. Thus there is a $\bZ_2$ classification. Meanwhile for $\bZ_4$, the symmetry class considered in Sections \ref{seclattice} and \ref{secfieldtheory}, there are no possible nontrivial charges.

The existence of a charged string with long-range order implies an exponentially localized edge mode. This is often phrased in terms of projective symmetry representations. Indeed, if we have a boundary, the symmetry will fractionalize on it as before, and now the charges of the $U_R$ define the commutation relations of a projective representation, which we identify with the ``anomalous" symmetry action at the edge. A more concrete way to see the edge degeneracy is to consider our system defined on an interval. The string operator stretching across the whole system satisfies
\[0 \neq \langle\cO^\dagger_0 (\prod_{0 \le x \le N} U_x) \cO_N\rangle = \langle \cO^\dagger_0 \cO_N\rangle,\]
where we used the global symmetry. If the end point operator is charged, it therefore acts as a symmetry breaking order parameter at the boundary.

In gapless systems, we still have symmetry fractionalization for gapped symmetries (i.e., symmetries which act non-trivially only on gapped degrees of freedom) and many of the previous arguments apply. For gapless symmetries (i.e. symmetries which act nontrivially on the gapless degrees of freedom) it does not hold. Nevertheless, one can still study string operators $(\prod_{x < n} U_x) \cO_n$, which will always have algebraic decay for a gapless symmetry. In \cite{verresen_gapless_2019} it was argued that the string operators with the slowest such decay define topological invariants for gapless phases, and can lead to edge modes, with either exponential or algebraic localization at the edge. In that setting, the string operators can actually be degenerate and have different charges, which complicates the bulk-boundary correspondence. We will see that for systems with an emergent anomaly, however, the charges are much more regular, since they have to satisfy the anomaly. Indeed, we will argue that we always have exponentially localized edge modes.

\subsection{String Operators in Anomalous Theories}\label{appstringopanoms}

Let us first consider 1d bosonic anomalies for a cyclic group $G = \bZ_n$. It is a consequence of the classification of anomalies that for any abelian group, the anomaly can be determined by its finite cyclic subgroups. For these, there is a $\bZ_n$ classification. Let $k \in \bZ_n$ be the level of the anomaly. In terms of the 2d SPT bulk, the gauge fluxes have fractional statistics with topological spin $\theta = 2\pi i k/n^2$ \cite{Levin_2012}. This also can be derived from the Chern-Simons form of the associated topological term \cite{kapustin2014anomalies}
\[\omega(A) = \frac{k}{n^2} A \cup dA.\]
See appendix \ref{appgaugefield}. Below we mention a direct connection between the $U(1)$ chiral anomaly and the $\bZ_n$ anomaly.

This topological spin translates into a certain spin-selection rule for the boundary \cite{chang_topological_2019,Lin_2019}. One finds that in the $g$-twisted sector, where $g$ is the generator of $\bZ_n$, all states have fractional spin (i.e. momentum around the circle)
\[\label{eqnspinselrule} S \in \frac{k}{n^2} + \frac{1}{n}\bZ.\]
Since the spin in the twisted sector is also the self-charge of the string operator, we find all string operators have \textit{fractional} self-charge with fractional part $k/n$. This spin-selection rule is both necessary and sufficient to diagnose the $\bZ_n$ anomaly. We note a possible confusion which is that the $\bZ_n$ charge of the bulk $\bZ_n$ flux (that is, in the 2d SPT) is $2k/n$ \cite{barkeshli_symmetry_2014}, which differs by a factor of two from the charge of the boundary string operator, see below.

We can see this spin-selection rule and fractional charge from the field theory of Section \ref{secfieldtheory}. The shift symmetry $\Phi_2 \mapsto \Phi_2 + \pi$ (resp. $\Theta_2 \mapsto \Theta_2 + \pi$) is generated by $\exp \frac{i}{2} \int_x \partial_x \Theta_2$ (resp. $\exp \frac{i}{2} \int_x \partial_x \Phi_2$). The string operators for these symmetries thus have the form

\[\exp\left(\frac{i}{2} \int_{-\infty}^x\partial_x \Theta_2\right)\cO_x=\exp\left(\frac{i}{2}\Theta_2(x)\right) \cO_x,\]
\[\exp\left(\frac{i}{2} \int_{-\infty}^x \partial_x \Phi_2\right) \cO_x' = \exp\left(\frac{i}{2} \Phi_2(x)\right) \cO_x',\]
respectively, where $\cO_x$ and $\cO_x'$ are local operators. Observe that these operators are not mutually local, but have a braiding phase of $\pm i$ (the two shift symmetries have a mutual anomaly). Thus, their fusion products, which are the string operators for the diagonal shift symmetry $R_x$, have spin $\pm 1/4$ mod 1, hence also fractional self-charge $\pm i$.

This highlights several recurring features of anomalous symmetries. First, unlike SPT phases where the charges of the string operators are ordinary linear charges, the charges of string operators in an anomalous theory are fractional or projective. This allows us to quickly see that our phases are not of the familiar SPT $\times$ gapless type.

Second, there is no nondegenerate, gapped symmetric phase, since by modularity the spin is related to the conformal dimension, which must thus be nonzero if the spin is nonzero. (In a fermionic system, the constraints of modular invariance are slightly relaxed, and allow for half-integer spins in a gapped phase, but the anomaly fractional spins which occur are always $\le 1/4$.) For even $n$, the fractional charge leads to degenerate string operators, which is impossible in a gapped phase (without symmetry breaking) \cite{pollmann_symmetry_2012}. For example if $n = 2$, then any string operator has charge $\pm i$ by the spin selection rule. Taking its Hermitian conjugate we get a degenerate string operator of charge $\mp i$.

This raises a third point, which is that all string operators in the anomalous theory are charged. Indeed, one can consider an anomaly to be an obstruction to gauging the symmetry \cite{Kapustin_2014,kapustin2014anomalies}. The fact that all string operators are charged means that there are no gauge-invariant states in the twisted sector of the gauge theory, which is a pathology. For edge modes, this means that the energy splitting is always exponentially small, with localization length set by the gap of the fundamental charges. We return to these edge modes in Appendix \ref{subappbulkboundary} below.

Several 1d anomalies can be understood in terms of the familiar chiral anomaly \cite{Lu_2012}. For a $U(1)$ symmetry there is a continuous family of twisted boundary conditions on a circle, which we think of as the magnetic flux of a $U(1)$ gauge field passing through it. As we vary the magnetic flux $\Phi$ there is a \textit{spectral flow}, where the energy levels of the system in the flux background move continuously as a function of $\Phi$ \cite{SCHWIMMER1987191}.

% The spectral flow is constrained to have a very simple form. If we begin with a state $|0\rangle$ of dimension $(h,\bar h)$ and $U(1)$ charge $(q,\bar q)$, we get a family of states $|\Phi\rangle$ of dimension
% \[h(\Phi) = h - \frac{\Phi}{2\pi} q + \frac{k \Phi^2}{8\pi^2}\]
% \[\bar h(\Phi) = \bar h + \frac{\Phi}{2\pi} \bar q + \frac{\bar k \Phi^2}{8\pi^2},\]
% where $k, \bar k \in \bZ$

% The holomorphic and antiholomorphic scaling exponents $(h_n,\bar h_n)$ of a state $|n\rangle$ of charge $q_n$
% \[h(\Phi) = h(0) - \Phi \]

The meaning of the chiral anomaly is that as $\Phi$ is taken from $0$ to $2\pi$, the spin of all operators shifts by an integer $k \in \bZ$, which is the level of the anomaly (for fermionic systems $k$ can be a half integer \cite{belov2005classification}). In a conformal field theory, if we begin with a neutral spin-zero state at $\Phi = 0$, the spins of the corresponding family of states along the spectral flow have the universal form \cite{Lin_2019}
\[S = \frac{k \Phi^2}{4 \pi^2}.\]
If $\Phi = 2\pi/n$, we can identify the ground states in the flux sector with the string operators for the $\bZ_n$ subgroup of $U(1)$ and we see the form of $S$ above matches our spin selection rule \eqref{eqnspinselrule} for the $\bZ_n$ anomaly.

The spectral flow above can be related to a bulk Hall current. As we increase the flux adiabatically from $0$ to $2\pi$, we end up producing charge. In a bulk+boundary setup, this charge is thought of as coming from the bulk (hence ``anomaly in-flow"). If we put our system on a cylinder with two circular boundaries, and thread a flux through the middle, the amount of charge pumped is identified with the bulk Hall current. Thus, one would like to identify the charge of the boundary twisted states with the charge of the bulk flux, but the charges of the former are half the charges of the latter. The way this is resolved in the bulk+boundary quantum Hall setup is that the bulk Chern-Simons term itself contributes to the boundary current \cite{BARDEEN1984421}, and makes up for the missing half of the charge, so the bulk Hall current corresponding to the above spectral flow is $2k$.

% Identifying the spin with the self-charge of the flux, we see the ``charge" bound to the $\Phi$-flux, in other words the charge inside the circle, is $k\Phi/2\pi$. Thus as $\Phi$ goes from $0$ to $2\pi$, $k$ units of charge flow across the circle. In a physical system with a local $U(1)$ charge, our circle must be the boundary of a 2d system, such as a cylinder. In this case we interpret flux threading the cylinder as inducing an adiabatic current which pumps $k$ units of charge from one boundary of the cylinder to the other, resulting in a quantized bulk Hall current.

Another important case for us is $U(1) \times \bZ_2$, which can be thought of as a subgroup of a mixed $U(1) \times U(1)$ anomaly such as between the vector and axial symmetries of a 1d compact boson. The mixed anomaly in this symmetry class pumps a $\bZ_2$ charge when we thread a $2\pi$ flux for the $U(1)$. Equivalently, we find that the $\bZ_2$ string carries half-integer charge under the $U(1)$.

Meanwhile, for $U(1) \rtimes \bZ_2^T$, the anomaly may be detected by the $U(1)$ $\pi$-flux string operator. Indeed, only a $0$ and $\pi$ flux have time reversal symmetry. The anomalous case is where the $\pi$ flux has a Kramers degeneracy \cite{kapustin2014bosonic}.

% Anomalies in higher dimensions have a similar flavor. For instance we will discuss the $SO(3) \times U(1)$ anomaly of the 2d deconfined quantum critical point. In this anomaly, the $U(1)$ flux, a string operator in 2d, carries a projective spin-1/2 representation of $SO(3)$ \cite{Wang_2017,Komargodski_2019, Metlitski_2018}.r

\section{Emergent Anomalies}\label{appemerganom}

\subsection{Anomaly Vanishing}\label{subappanomvanishing}

In this appendix we describe the anomaly vanishing equation which must be satisfied by an emergent anomaly which appears in a system with a microscopic on-site symmetry. We will show how the equation is solved in our $\bZ_4$ example as well as other symmetry classes which have appeared in the literature, namely $U(1) \times \bZ_2$ and $U(1) \rtimes T$. We also describe how anomaly vanishing applies to systems with Lieb-Schultz-Mattis constraints and the 2d deconfined quantum critical point. We end with a discussion where we argue that any global anomaly may be realized as an emergent anomaly in a system with on-site symmetry action. Some details on discrete gauge fields may be found in Appendix \ref{appgaugefield}.

Let $G$ be the microscopic symmetry group, $G_{\rm low}$ be the quotient of $G$ which is realized on the low energy degrees of freedom, and $G_{\rm gap}$ the (normal) subgroup of $G$ which acts only on the gapped degrees of freedom. We have $G_{\rm low} = G/G_{\rm gap}$. Let $\pi:G \to G_{\rm low}$ be the quotient map. One considers the anomaly as classified by an SPT phase in one higher dimension, which is in turn associated with an element $[\omega]$ in a cohomology theory such as group cohomology \cite{Chen_2013} or spin cobordism \cite{kapustin2014symmetry,freed2016reflection}. There is an associated map $\pi^*$ from the group of $G_{\rm low}$ SPT phases to $G$ SPT phases. Since in a system with on-site microscopic symmetry, there is no anomaly, we must have
\[\label{eqnanommatching}\pi^* [\omega] = 0,\]
which we refer to as the anomaly vanishing equation. See Appendix \ref{appgaugefield} for a proof.

For the examples we study, it suffices to consider this equation in group cohomology, where $[\omega]$ is represented by some group cocycle $\omega(A_{\rm low})$, which can be thought of as the effective action of the higher dimensional SPT controlling the anomaly, coupled to a background $G_{\rm low}$ gauge field $A_{\rm low}$ \cite{kapustin2014anomalies}. The anomaly vanishing equation may be rewritten
\[\label{eqnanommatchingalt}\exists \alpha \qquad \omega(A_{\rm low}) = d \alpha(A_{\rm low},A_{\rm gap}),\]
where $\alpha(A_{\rm low},A_{\rm gap})$ can be thought of as a boundary counterterm involving background gauge fields $A_{\rm low}$ for $G_{\rm low}$ \textit{and} $A_{\rm gap}$ for $G_{\rm gap}$. We will see that this counterterm has a physical interpretation that allows us to reason about edge modes. In a way it is like the topological term for the background gauge fields generated by integrating out the gapped degrees of freedom. Indeed, if $Z(A_{\rm low})$ is the partition function of the gapless theory coupled to background $G_{\rm low}$ gauge field, then although this is not gauge invariant, because of the anomaly, the anomaly vanishing equation is equivalent to saying that
\[\label{eqnanompartfun}Z_{\rm low}(A_{\rm low}) e^{2\pi i \int \alpha(A_{\rm low},A_{\rm gap})}\]
is gauge invariant when $(A_{\rm low},A_{\rm gap})$ is interpreted as a $G$ gauge field. Up to terms exponentially small in the gap, this is the partition function of the full theory.

The background gauge field $A_{\rm low}$ is a usual background gauge field (see appendix \ref{appgaugefield} for a review), and satisfies 
\[\label{eqnlowcocycle}dA_{\rm low} = 0\]
when $G_{\rm low}$ is discrete, but when the group extension
\[G_{\rm gap} \to G \to G_{\rm low}\]
is nontrivial, then $A_{\rm gap}$ sees a flux background defined by $A_{\rm low}$, meaning
\[\label{eqngappedgaugefield}dA_{\rm gap} = c(A_{\rm low}),\]
where $c \in H^2(BG_{\rm low},Z(G_{\rm gap}))$ is a 2-cocycle associated with the group extension and $Z(G_{\rm gap})$ is the center of $G_{\rm gap}$ (see below) \cite{brown2012cohomology}. Note that if $G$ is not a central extension, then $G_{\rm low}$ acts on $G_{\rm gap}$ and this cohomology must be considered to be twisted by this action. See \cite{Thorngrenthesis} for an introduction. Likewise, in expressions such as $dA_{\rm gap} = c(A_{\rm low})$, $dA_{\rm gap}$ must be considered the $A_{\rm low}$-twisted differential of $A_{\rm gap}$. Even when $c = 0$, this can lead to solutions to the anomaly vanishing equation \cite{splitextensions}. However, in the case of central extensions, which includes all the examples in this work, if $c = 0$, then there are no nontrivial solutions to the anomaly vanishing equation for bosonic extensions (although see Appendix \ref{subsubappfermtrivs} below).

To derive \eqref{eqngappedgaugefield}, we write the $G$ gauge field as
\[A = j(A_{\rm gap}) + s(A_{\rm low}),\]
where $j:G_{\rm gap} \to G$ is the inclusion map, and $s:G_{\rm low} \to G$ is a section (not necessarily a group homomorphism) of the quotient $\pi:G \to G_{\rm low}$, meaning $\pi \circ s$ is the identity on $G_{\rm low}$. We have
\[dA = 0 \qquad \Leftrightarrow \qquad j(dA_{\rm gap}) + ds(A_{\rm low}) = 0.\]
The extension cocycle is defined by
\[j(c(A_{\rm low})) = - ds(A_{\rm low}).\]
It is independent of $s$ up to gauge transformations. Thus, we find \eqref{eqngappedgaugefield}. Since a gauge field $A$ on a manifold $X$ can be interpreted as an element in $Z^1(X,G)$, we see that $c: Z^1(X,G_{\rm low}) \to Z^2(X,Z(G_{\rm gap}))$. By the classification of such maps discussed in Appendix~\ref{appgaugefield}, we can identify $c$ with an element in $H^2(BG_{\rm low},Z(G_{\rm gap}))$, as claimed above.

Let us note that when the system is completely gapped, so $G = G_{\rm gap}$, then necessarily $\omega = 0$, so the anomaly vanishing equation is simply
\[d\alpha = 0.\]
In this case we identify $\alpha$ with the SPT cocycle which characterizes this gapped phase. This illustrates that $\alpha$ is indeed part of the physical data which defines the 1d phase. For a given $\omega$, solutions to the anomaly vanishing equation form a torsor over the $G$ SPT classes. Physically this means that different solutions to the anomaly vanishing equation correspond with stacking an SPT phase on top of our system.

% We summarize this as follows:

% \begin{shaded}
% \textit{\textbf{Summary.} For a given microscopic on-site symmetry $G$ and effective low energy symmetry $G_{\rm low}$, possible intrinsically gapless $G$-SPT phases are labeled by elements of the subgroup
% \begin{equation}
% {\rm Ker} \left( \pi^* \right) \subset H^3(BG_{\rm low},U(1))
% \end{equation}
% where the homomorphism $\pi: G \to G_{\rm low}$ describes how the microscopic symmetries are realized in the effective low energy theory, with ${\rm Ker}(\pi) = G_{\rm gap}$, and $\pi^*$ is the induced map on cohomology.}
% \end{shaded}

% \begin{shaded}
% \textit{For a given microscopic on-site symmetry $G$ and gapped symmetry $G_{\rm gap}$, possible intrinsically gapless $G$-SPT phases are labeled by elements of the subgroup
% \begin{equation}
% {\rm Ker} \left( \pi^* \right) \subset H^3(B(G/G_{\rm gap}),U(1))
% \end{equation}
% where the homomorphism $\pi^*$ is the map on cohomology induced by the quotient.\newline
% \indent When the effective low energy symmetry $G_{\rm low}$ is larger than $G/G_{\rm gap}$ (which it contains as a subgroup), we restrict to elements of the kernel which extend to $G_{\rm low}$.}
% \end{shaded}

% To summarize, with a microscopic on-site symmetry $G$, intrinsically gapless topological phases are possible for each triple $(G_{\rm low},\pi,\omega)$ of a low energy symmetry $G_{\rm low}$, a map $\pi:G \to G_{\rm low}$, and an emergent anomaly $\omega$ for the low energy symmetry, such that
% \[\label{eqnanomvanishing}\pi^* \omega = 0.\]

\subsection{Examples}\label{subappexamples}

Let us show \eqref{eqnanommatching} holds for the symmetries and anomalies we have considered.

% We describe these gauge fields in terms of 1-cocycles $A_{\rm low} \in Z^1(X,G_{\rm low})$ and $A_{\rm gap} \in C^1(X,G_{\rm gap})$, where $X$ is a spacetime. See the appendix of \cite{thorngren2020tqft} for a review. We can think of the quotient as a group extension
% \[G_{\rm gap} \to G \to G_{\rm low}.\]
% There is an associated 2-cocycle $c \in Z^2(BG_{\rm low},G_{\rm gap})$, meaning $c(A_{\rm low}) \in Z^2(X,G_{\rm gap})$. This cocycle modifies the flatness condition for the $G_{\rm gap}$ gauge field to
% This equation will be the key to solving \eqref{eqnanommatching}. The way the pullback map $\pi^*$ works is that the cohomology class of $\omega$ is defined by some formula of $A_{\rm low}$ up to exact pieces made from $A_{\rm low}$, while $\pi^* \omega$ is defined by that same formula but its cohomology class is defined modulo exact pieces made out of both $A_{\rm low}$ and $A_{\rm gap}$, satisfying the above.

\subsubsection{$G = \bZ_4$}

In the case $G = \bZ_4$, $G_{\rm gap} = \bZ_2$, $G_{\rm low} = \bZ_2$, working in group cohomology, the anomaly can be written in the Chern-Simons-like form \cite{kapustin2014anomalies}
\[\omega = \frac{1}{2} A_{\rm low} \cup \frac{dA_{\rm low}}{2}.\]
Note that $dA_{\rm low} = 0$ mod 2, so $\frac{dA_{\rm low}}{2}$ is an integer class (it is equivalent to the Bockstein of $A_{\rm low}$). Meanwhile
\[\label{eqnZ4extension}c(A_{\rm low}) = \frac{d A_{\rm low}}{2},\]
so we can write
\[\label{eqnZ4counterterm}\pi^* \omega = d\left( \frac{1}{2} A_{\rm low} \cup A_{\rm gap} \right),\]
which means \eqref{eqnanommatching} is satisfied in cohomology. Indeed, using the product rule, \eqref{eqnlowcocycle}, \eqref{eqngappedgaugefield}, and \eqref{eqnZ4extension}, we have
\[d\left( \frac{1}{2} A_{\rm low} \cup A_{\rm gap} \right) = \frac{1}{2} dA_{\rm low} \cup A_{\rm gap} + \frac{1}{2} A_{\rm low} \cup dA_{\rm gap} = \frac{1}{2} A_{\rm low} \cup c(A_{\rm low}) = \frac{1}{2} A_{\rm low} \cup \frac{dA_{\rm low}}{2}.\]
See \cite{Wang_2016} for another perspective on the trivialization of this 2d SPT class.

The meaning of the term in parentheses is that the $G_{\rm low}$ string is charged under $G_{\rm gap}$, indeed as it must be to match the fractional charge. We return to this point in Appendix \ref{subappbulkboundary} below.

\subsubsection{$G = \bZ_2 \times U(1)$}\label{appU1timesZ2}

The case with $G = \bZ_2 \times U(1)$, $G_{\rm gap} = \bZ_2$ generated by the order two element of $U(1)$ is particularly common in the literature, especially where the $U(1)$ is particle number, so $G_{\rm gap}$ is the fermion parity, meaning we are in a gapless phase where the fermion is gapped. For such models we write $A_{\rm low}$ as a pair of a $U(1) = U(1)/\bZ_2$ gauge field $A_{\rm low}^{U(1)}$ and a $\bZ_2$ gauge field $A_{\rm low}^{\bZ_2}$. The extension class is the first Chern class
\[c(A_{\rm low}) = \frac{dA_{\rm low}^{U(1)}}{2\pi}.\]
The effective anomaly is
\[\label{eqnanomU(1)Z2}\omega = \frac{1}{2} A_{\rm low}^{\bZ_2} \cup \frac{dA_{\rm low}^{U(1)}}{2\pi}.\]
This anomaly is realized for instance via the 1d chiral anomaly, where $A_{\rm low}^{U(1)}$ couples to the vector current and $A_{\rm low}^{\bZ_2}$ to the axial current. When we consider \eqref{eqngappedgaugefield} however, we find
\[\label{eqnU(1)Z2counterterm}\pi^* \omega = d\left( \frac{1}{2} A_{\rm low}^{\bZ_2} \cup A_{\rm gap} \right).\]
When $G$ is broken to $\bZ_2 \times \bZ_2$, the term in parentheses is a nontrivial SPT class, so the topology in such gapless phases is essentially due to an SPT sector. Extending the parity symmetry to $U(1)$, however, requires a vanishing gap because of the anomaly \eqref{eqnanomU(1)Z2}. We note that if one considers the $\bZ_4$ subgroup generated by the generator of the $U(1)$ times the generator of the $\bZ_2$, then such examples reduce to the calculation above, and in this symmetry class the topological phase is intrinsically gapless.

\subsubsection{$G = U(1) \rtimes T$}\label{appU1timest}

The case $G_{\rm gap} = \bZ_2$, $G_{\rm low} = U(1) \rtimes \bZ_2^T$ is also quite interesting, although similar. Because of the time reversal symmetry, we must work with the first Stiefel-Whitney class $w_1 \in Z^1(X,\bZ_2)$, which plays the role of the time-reversal gauge field \cite{kapustin2014symmetry}. We also must treat the $U(1)$ background $A_{\rm low}$ as a background $U(1)$ gauge field with curvature $F_{\rm low}$. The formula for the emergent anomaly can be found in \cite{kapustin2014bosonic}. It is
\[\omega = \frac{1}{2} w_1 \cup \frac{F_{\rm low}}{2\pi}.\]
The physics of this term is that the $\pi$ flux of $U(1)_{\rm low}$ carries a Kramers doublet under $T$.

Since both the low energy $U(1)$ and time reversal symmetries are extended by fermion parity (to a group called $Pin(2)^-$), the extension class may be written
\[c(A_{\rm low},w_1) = w_1 \cup w_1 + \frac{F_{\rm low}}{2\pi}.\]
We see
\[\label{eqnU1Tcounterterm}d\left(w_1 \cup \frac{1}{2}A_{\rm gap}\right) = \frac{1}{2} w_1 \cup w_1 \cup w_1 + \pi^*\omega.\]
The first term is actually exact by a Wu relation \cite{thorngren2020tqft}. Thus, we find \eqref{eqnanommatching} is satisfied.

The meaning of the term in parentheses is that the $\pi$ flux of $U(1)_{\rm low}$ carries odd fermion parity, indeed as it must be to match the Kramers degeneracy. We return to this point in Appendix \ref{subappbulkboundary} below.

\subsubsection{2+1D Deconfined Quantum Critical Point}\label{appdqcpanomcalc}

These calculations also apply to higher dimensions, as we demonstrate in the example of the $SO(3) \times U(1)$ anomaly of the $\mathbb{CP}^1$ model, associated to various deconfined quantum critical points (DQCPs). The anomaly is \cite{Wang_2017,Komargodski_2019,Metlitski_2018}
\[\omega = \frac{1}{2} w_2(A_{\rm low}^{SO(3)})\cup \frac{dA_{\rm low}^{U(1)}}{2\pi},\]
where $A_{\rm low}^{SO(3)}$ is the background $SO(3)$ gauge field, $w_2$ is its 2nd Stiefel-Whitney class, which obstructs lifting it to an $SU(2)$ gauge field, and $A_{\rm low}^{U(1)}$ is the background $U(1)$ gauge field. If fundamental fermions are gapped where the $SO(3)$ is spin rotation and the $U(1)$ is particle number, then both groups get extended by $G_{\rm gap} = \bZ_2$ according to
\[c(A_{\rm low}^{SO(3)},A_{\rm low}^{U(1)}) = w_2(A_{\rm low}^{SO(3)}) + \frac{dA_{\rm low}^{U(1)}}{2\pi}.\]
We find
\[\label{eqnappdqcpanomcanc}d\left( \frac{1}{2} A_{\rm gap}\cup \frac{dA_{\rm low}^{U(1)}}{2\pi} + \frac{1}{4\pi} A_{\rm low}^{U(1)} \wedge \frac{dA_{\rm low}^{U(1)}}{2\pi}\right) = \pi^*\omega.\]
We note the second term on the left-hand side is a \textit{half-quantized} $U(1)_{\rm low}$ Chern-Simons term (since the low energy theory is bosonic).

% i.e. a $U(1)$ Chern-Simons term of level $2$ mod $4$, and is exact, so $\pi^* \omega$ is trivial in cohomology, as required for such an anomaly to emergent from an on-site symmetry $(SU(2) \times U(1))/\bZ_2$ when the fermion is gapped. The interpretation of the term in parentheses is that the $2\pi$ flux of $U(1)_{\rm low}$ carries odd fermion parity.

Another way we can cure the anomaly is by considering
\[\label{eqnappdqcpanomcanc2}d\left( \frac{1}{2} A_{\rm gap}\cup w_2(A_{\rm low}^{SO(3)}) + \frac{1}{2} CS(A_{\rm low}^{SO(3)})\right) = \pi^*\omega,\]
where $CS(A_{\rm low}^{SO(3)})$ is the $SO(3)$ Chern-Simons term of smallest level (it looks like level 4 for $SU(2)$), which satisfies
\[ \int dCS(A_{\rm low}^{SO(3)}) = \int w_2(A_{\rm low}^{SO(3)}) \cup w_2(A_{\rm low}^{SO(3)}) \mod 2.\]
We see the two solutions to the anomaly vanishing equation differ by a Chern-Simons term for the full group $G = Spin^c(3)$.

A similar result holds for the largest possible unitary symmetry $SO(5)$ of the DQCP. In this case the emergent anomaly is
\[\omega = \frac{1}{2} w_4(A_{\rm low}^{SO(5)}),\]
where $w_4$ is the 4th Stiefel-Whitney class. The group extension by fermion parity $G_{\rm gap} = \bZ_2$ is
\[\bZ_2 \to Spin(5) \to SO(4),\]
again classified by the 2nd Stiefel-Whitney class. The integer cohomology of $BSpin(5)$ is torsion-free (for instance, using the isomorphism $Spin(5) = Sp(2)$), so since $\omega$ is associated with a 2-torsion class, $\pi^* \omega = 0$ in cohomology. This cancellation underlies the fermionic model for DQCP with on-site $Spin(5)$ symmetry in \cite{MikeZilla}. By naturality of the pullback, the anomaly is therefore microscopically trivial (admitting an on-site representation) for any subgroup $G \subset Spin(5)$.

% The second term in \eqref{eqnappdqcpanomcanc} is a $\theta = \pi$ term for the low energy $U(1)$. In the presence of time reversal symmetry, it is not exact.

\subsubsection{Fermionic Trivializations}\label{subsubappfermtrivs}

In the case of fermionic systems, because fermionic SPTs are classified by spin cobordism, occasionally the group cohomology incarnation of the anomaly vanishing equation we used above is insufficient. This can lead to some new mechanisms for anomaly vanishing which work even when the group extension is trivial, so $c = 0$.

For instance, on spin manifolds, the 2nd Stiefel-Whitney class of the tangent bundle is exact: $w_2 = d\eta$. In some sense $\eta$ can be thought of as the spin structure \cite{Thorngren_2015,Thorngrenthesis}. Some bosonic anomalies have an $\omega$ which is ``proportional" to $w_2$ and in this case we can write
\[\omega(A_{\rm low}) = d\alpha(A_{\rm low},\eta).\]

For example, there is time-reversal-protected 2+1D anomaly/3+1D SPT with
\[\omega = \frac{1}{2} w_1^4 = \frac{1}{2} w_2 w_1^2\]
(known as $eTmT$ in the notation of \cite{WangPotterSenthil}), where we have used the Wu formula. This becomes trivial with the addition of $T^2 = 1$ fermions because
\[\omega = d\left(\frac{1}{2} \eta w_1^2\right).\]
This anomaly is not cancellable with $T^2 = (-1)^F$ fermions however since in that case $d\eta = w_2 + w_1^2$ \cite{thorngren2018anomalies}.

An example which does not involve time reversal is that of a $\bZ_2$ 1-form symmetry anomaly in 2+1D/SPT in 3+1D with
\[\omega = \frac{1}{2} B^2 = \frac{1}{2} B w_2,\]
where $B$ is a background $\bZ_2$ 2-form gauge field. This anomaly is again cancelled by adding fermions:
\[\omega = d\left( \frac{1}{2} B \eta \right).\]
This particular anomaly vanishing actually underlies bosonization in 2+1D \cite{thorngren2018anomalies,Gaiotto_2016}.

\subsection{Solvability of the Anomaly Vanishing Equation}\label{subappsolvability}

Let us discuss when the anomaly vanishing equation \eqref{eqnanommatching} can be solved in group cohomology. For a given $G_{\rm low}$ and $\omega$, there is usually a $G$ and $\pi$ where $\omega$ is in ${\rm Ker}(\pi^*)$. For group cohomology SPTs, this is always the case. In \cite{Tachikawa_2020,Wang_2018} the authors considered a technique for constructing \textit{gapped} boundaries of SPTs, where $G_{\rm gap}$ plays the role of a gauge symmetry, and $G_{\rm low}$ that of the microscopic symmetry, where the same mathematical problem appears\footnote{Our constructions can be used to create anomalous theories with microscopic $G_{\rm low}$ symmetry by gauging $G_{\rm gap}$, but of course they won't be gapped.}. They showed that for any finite $G_{\rm low}$, there is a corresponding extension $G$ for which the SPT may be disentangled in this way.

Applied to our setting, these results imply that any system with finite anomalous $G_{\rm low}$ symmetry is realized in a system with some on-site $G$, of which $G_{\rm low}$ is a quotient. Using the results of \cite{thorngren2020tqft} (where one of us studied gapped boundaries for SPTs protected by Lie groups), we can extend this to the case of \textit{global} anomalies of continuous symmetry groups, meaning we have to exclude anomalies such as the chiral anomaly, which can be diagnosed by local correlation functions, for which no gapped charges can cure the anomaly.

For fermion SPTs/anomalies beyond group cohomology, the conditions are more subtle. For example, a unitary symmetry $\bZ_2 \times \bZ_2^F$ protects a $\bZ_8$ classification of fermion SPTs in 2d, or equivalently anomalies in 1d. The generators of this group cannot be trivialized by any extension of the $\bZ_2$ symmetry (fermion parity cannot be extended). Indeed, when $\bZ_2$ is extended to $\bZ$, a $\bZ_2$ classification remains, where the domain wall carries a Kitaev wire. Another example is the $\bZ_{16}$ classification of $T^2 = (-1)^F$ fermion SPT phases in 3d. Again under symmetry extension at most we can reduce $\bZ_{16} \to \bZ_2$, since the domain wall on the boundary will always carry a chiral mode of $c_- = 1/2$.

\subsection{Construction of Instrinsically Gapless SPTs}\label{subappconstruction}

\begin{figure}
    \centering
    \includegraphics[width=6cm]{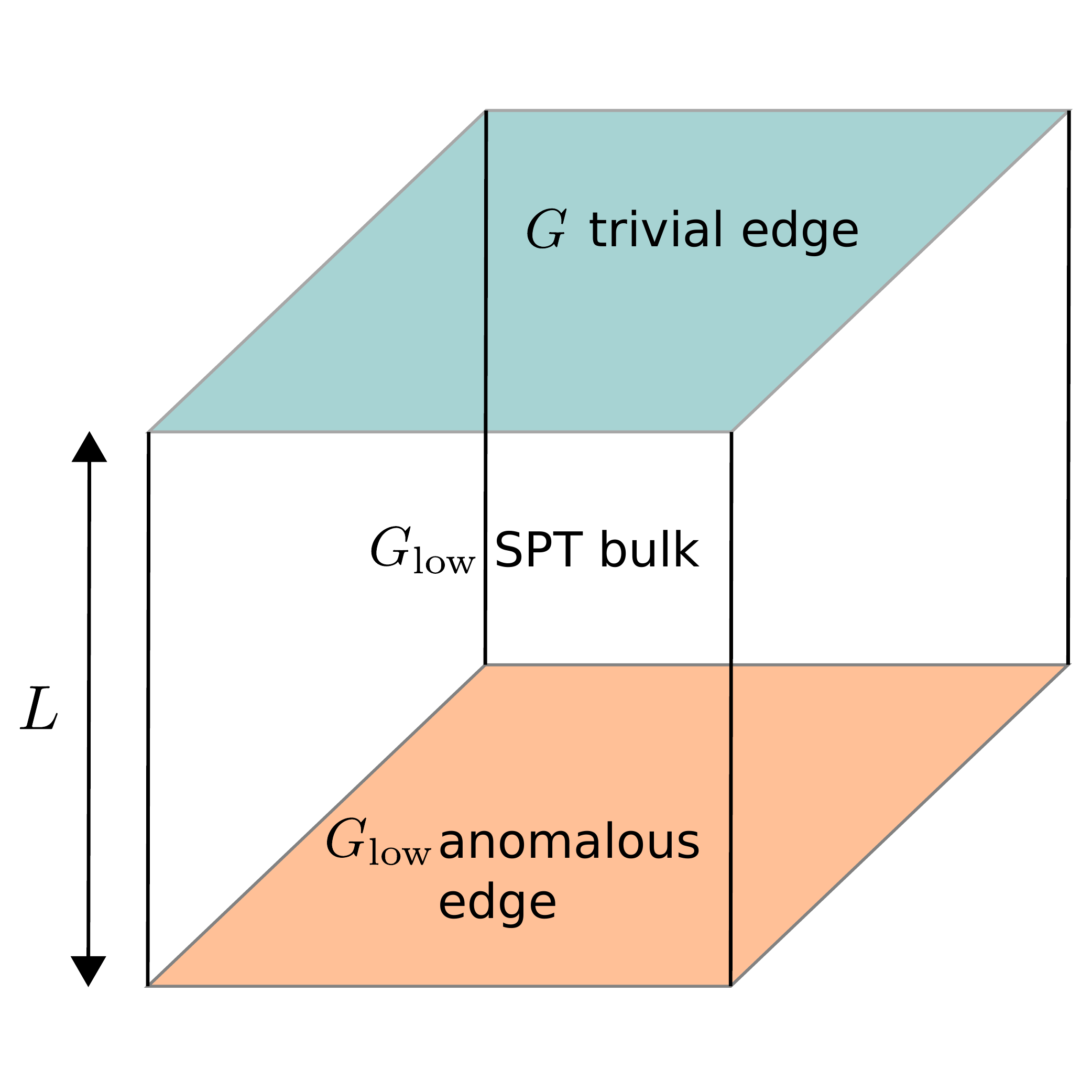}
    \caption{The ``slab construction" of intrinsically gapless SPT phases. The $G$ symmetry acts on-site everywhere, but inside the SPT bulk and along the anomalous edge it acts by the quotient $G \to G_{\rm low}$. In terms of the partition function \eqref{eqnpartfun}, the bottom edge contributes $Z_{\rm low}(X,A_{\rm low})$ and the top edge contributes $e^{2\pi i \int_X \alpha(A)}$. To define the purely $d$-dimensional system, we keep the width of the slab $L$ fixed and take the horizontal directions to be infinite.}
    \label{fig:slab}
\end{figure}

Let us discuss how to construct intrinsically gapless SPT phases. From the point of view of field theory, the topological term $\alpha$ we get from solving the anomaly vanishing equation is enough to give an effective action for this phase, but what if we want a lattice model?

Suppose we begin with a $d$-space-dimensional lattice model for the anomalous $G_{\rm low}$ theory as a boundary of a $d+1$-dimensional $G_{\rm low}$ SPT, so that $G_{\rm low}$ acts on-site in the whole system. Since we want to construct a $d$-dimensional system, we suppose that the bulk SPT has only some finite width $L$, forming a slab geometry $Y_d \times [0,L]$, with our anomalous theory localized to $Y_d \times 0$. See Fig. \ref{fig:slab}.

On the other boundary $Y_d \times L$ we place microscopic degrees of freedom transforming in an on-site representation of $G$. Because of the nontrivial extension, these degrees of freedom look projective from the point of view of $G_{\rm low}$. The anomaly vanishing equation tells us that we can create a featureless edge along $Y_d \times L$ without breaking the $G$ symmetry (which acts through its quotient $G \to G_{\rm low}$ in the bulk and along $Y_d \times 0$) (compare \cite{Tachikawa_2020,Wang_2018}). Different featureless edges will correspond to different solutions of the anomaly vanishing equation by layering $d$-dimensional $G$ SPTs along $Y_d \times L$. See Appendix \ref{subapptrivialization} for a general construction for finite $G$.

Once we have ``trivialized" the $Y_d \times L$ edge we may be worried that we have gapped out the anomalous $Y_d \times 0$ edge. However, this coupling can at most introduce local $G_{\rm low}$-symmetric perturbations of the low energy modes, and these can be cancelled by tuning parameters along the $Y_d \times 0$ edge. These perturbations can be controlled by taking $L$ much larger than the bulk correlation length.

In the end, we produce a system on the slab $Y_d \times [0,L]$ with global on-site $G$ symmetry, and low energy theory described by our starting $G_{\rm low}$-anomalous theory. Since $L$ remains fixed, this is a $d$-dimensional system (although one which has a very large local Hilbert space). It shows that to any system with $G_{\rm low}$ anomaly and a solution to the anomaly vanishing equation, so long as we can construct a lattice model for this anomalous system, we can construct a lattice model for the associated intrinsically gapless SPT.

% From the point of view of the bulk-boundary correspondence, the $G_{\rm low}$ anomaly in $d$ dimensions is captured by an SPT in $d+1$ dimensions.

% This anomaly can be realized in a system with on-site $G$ symmetry precisely when it is possible to symmetrically disentangle this SPT by adding product state degrees of freedom transforming in representations of $G$ which are projective with respect to $G_{\rm low}$. 
% If one begins with the $d+1$-dimensional $G_{\rm low}$ SPT on a half-space, with our anomalous system at the boundary, and applies this disentangling procedure to the bulk, one obtains a realization of this anomalous system with an on-site $G$ symmetry. Thus, as long as one can realize a gapless symmetric boundary of the $G_{\rm low}$ SPT, then this procedure will construct an intrinsically gapless $G$-SPT phase.

\subsection{Trivialization of the SPT Boundary in a Fixed Point Model}\label{subapptrivialization}

To show how a solution to the anomaly vanishing equation $\omega(A_{\rm low}) = d\alpha(A)$ allows us to define a trivial $G$-symmetric edge of the $G_{\rm low}$ SPT classified by $\omega$, we can adapt a construction of \cite{Chen_2013} for finite $G$. First let us recall that construction. We will take our local Hilbert space to be associated with the vertices of a triangulated spatial manifold $Y$, with the local Hilbert space at a single vertex spanned by basis vectors labelled by elements of $G_{\rm low}$. $G_{\rm low}$ acts on these vectors in the so-called regular representation, i.e. by multiplication (and later $G$ will act by its quotient $G \to G_{\rm low}$). Thus, the whole Hilbert space is spanned by states $|\phi\rangle$ associated to $G_{\rm low}$-valued 0-cochains $\phi \in C^0(Y,G_{\rm low})$ (see Appendix \ref{appgaugefield} for notation and concepts used below).

We define the operator $Z_y^g$ near a vertex $y$ that multiplies the $G_{\rm low}$ label at $y$ by $g$ and also produces the following phase factor
\[e^{i\beta(\phi,y,g)} = \exp\left(2\pi i \int_{Y \times [0,1]} \omega(A(\phi,y,g))\right),\]
where $A(\phi,y,g)$ is a gauge field on the ``spacetime" $Y \times [0,1]$ which restricts to $1^\phi$ on $Y \times 0$ and $1^{\phi'}$, where $\phi'$ is the new $\phi'$ on $Y \times 1$. Thus, along the ``vertical" edge connecting $y \times 0$ to $y \times 1$, $A$ has label $g$, while on other ``vertical edges" (i.e. those connecting $Y \times 0$ and $Y \times 1$) it has the identity label. Thus, the phase factor $\beta(\phi,y,g)$ although we have expressed it as an integral over all of $Y \times [0,1]$, only receives nontrivial contributions from simplices touching $y$, so $Z_y^g$ so defined is a local operator. Writing it as an integral over all of $Y \times [0,1]$ however makes it clear that since $\omega$ is gauge invariant (by virtue of $d\omega = 0$), the $Z_y^g$'s satisfy the group algebra at $y$ and commute at separated points. Thus, the Hamiltonian
\[H = -\sum_y \left(\frac{1}{|G_{\rm low}|} \sum_g Z_y^g\right)\]
is a commuting projector Hamiltonian. Its unique ground state on a closed $Y$ may be written
\[|\omega \rangle = \sum_\phi e^{2\pi i \int_Y \nu(\phi)} |\phi\rangle,\]
where
\[d\nu(\phi) = \omega(1^\phi).\]
This reproduces the SPT ground state in \cite{Chen_2013} which corresponds to $\omega$.

Now suppose we have a solution to the anomaly vanishing equation $\omega(A_{\rm low}) = d\alpha(A)$. We define a Hilbert space on a spatial manifold $Y$ with boundary in much the same way as above, but now vertices along $\partial Y$ have Hilbert spaces spanned by elements of the bigger group $G$. We can write this as a combination of a $G_{\rm low}$-valued 0-cochain $\phi_{\rm low} \in C^0(Y,G_{\rm low})$ and a $G$-valued 0-cochain $\phi \in C^0(\partial Y, G)$, satisfying the condition $\phi_{\rm low}|_{\partial Y} = \pi(\phi)$, where $\pi:G \to G_{\rm low}$ is the quotient map.

We define $Z_y^g$ to be associated with a group element $g \in G$. On general vertices of $Y$, it acts by multiplication on $\phi_{\rm low}$ by the image of $g$ in $G_{\rm low}$, while on boundary vertices it also acts by multiplication in $G$ on $\phi$. We see it preserves the boundary condition $\phi_{\rm low}|_{\partial Y} = \pi(\phi)$. The associated phase factor is now
\[e^{i\beta(\phi,y,g)} = \exp\left(2\pi i \int_{Y \times [0,1]} \omega(A_{\rm low}(\phi_{\rm low},y,g)) - 2\pi i \int_{\partial Y \times [0,1]} \alpha(A(\phi,y,g))\right),\]
where $A_{\rm low}(\phi_{\rm low},y,g)$ is a $G_{\rm low}$ gauge field on $Y \times [0,1]$ defined as above to restrict to $1^{\phi_{\rm low}}$ on $Y \times 0$ and $1^{\phi_{\rm low}'}$ on $Y \times 1$, where $\phi_{\rm low}'$ is the transformed $\phi_{\rm low}'$. Likewise, $A(\phi,y,g)$ is a $G$ gauge field on $\partial Y \times [0,1]$ which restricts to $1^\phi$ on $\partial Y \times 0$ and to $1^{\phi'}$ on $\partial Y \times 1$. This phase factor reduces to local contributions from simplices touching $y$ and by the anomaly vanishing equation, which implies gauge invariance of this integral so long as $\phi_{\rm low}$ and $\phi$ are fixed along $Y \times 0$ and $Y \times 1$, the $Z_y^g$ so defined satisfy the group algebra at $y$ and commute at separated points. Thus, we can define the commuting projector Hamiltonian
\[H = -\sum_y \left(\frac{1}{|G|} \sum_g Z_y^g\right),\]
which has a unique $G$-symmetric ground state.

We expect a similar construction for supercohomology phases can be adapted from \cite{Kobayashi_2019,chen2020disentangling}. It would be very interesting to understand such constructions for continuous symmetry groups.

\subsection{Bulk-Boundary Correspondence}\label{subappbulkboundary}

We have seen in the above examples that in 1d the emergent anomaly is often associated with a fractional charge or projective symmetry representation of a string operator for $G_{\rm low}$. From the perspective of the full microscopic symmetry $G$ however, these fractional charges or projective representations are actually integral or linear. This necessitates that certain $G_{\rm low}$ string operators are charged under $G_{\rm gap}$. We will argue this and show it leads to edge modes.

Let $g \in G$ be a gapless symmetry. Observe that all $g$ algebraic string operators must have the same $G_{\rm gap}$ charge (which thus must be abelian). Otherwise, by fusion of these string operators, we would obtain a local algebraic operator with non-trivial $G_{\rm gap}$ charge, contradicting the definition of $G_{\rm gap}$.

We will show that if $h \in G_{\rm gap}$ acts nontrivially in any of these twisted sectors, that the unique $h$-string with long-range order has nontrivial $G$ charge. In a certain sense we will discuss, at any symmetric boundary condition this string operator has a vev. Since it defines a local order parameter at the boundary, there must be spontaneous symmetry breaking at the edge.

Let us be more precise. Suppose algebraic $g$ strings have charge $\chi_g:G_{\rm gap} \to U(1)$. This charge is multiplicative in $g$, so it defines a bicharacter (a homomorphism) $\chi:G \times G_{\rm gap} \to U(1)$. We will argue that for $h \in G_{\rm gap}$, the $h$ string with long-range order has $G$ charge $\chi(-,h):G \to U(1)$. This charge is uniquely defined because $G$ is unbroken.

The $g \in G$ charge of the $h \in G_{\rm gap}$ string operator with longest range order is captured by the leading term of the partition function on a rectangular torus of aspect ratio $\beta$, with a $g$ twist in the time direction and an $h$ twist in the space direction, in the ``low temperature" limit $\beta \to \infty$. We denote this partition function
\[Z(h/g,\beta) = {\rm Tr}_{\mathcal{H}_h} g e^{-\beta H},\]
where $\mathcal{H}_h$ is the Hilbert space in the $h$-twisted sector and $H$ is the Hamiltonian. In a conformal invariant IR limit, if we perform a modular $S$ transformation, we will find
\[Z(h/g,\beta) = Z(g/h,1/\beta) = {\rm Tr}_{\mathcal{H}_g} h e^{- H/\beta}.\]
Because all $g$-twisted states of the CFT limit have the same $G_{\rm gap}$ charge, in this case $\chi(g,h)$, this partition function is
\[e^{i \chi(g,h)} Z_{\rm low}(1/\beta,g/1),\]
where $Z_{\rm low}(1/\beta,g/1)$ is positive for all $\beta$. Thus, the phase of the leading term in the limit $\beta \to \infty$ of $Z(h/g,\beta)$ is $e^{i \chi(g,h)}$, proving the claim of reciprocity, at least in the CFT case. We believe this claim holds in full generality.

\begin{figure}
    \centering
    \includegraphics[width=6cm]{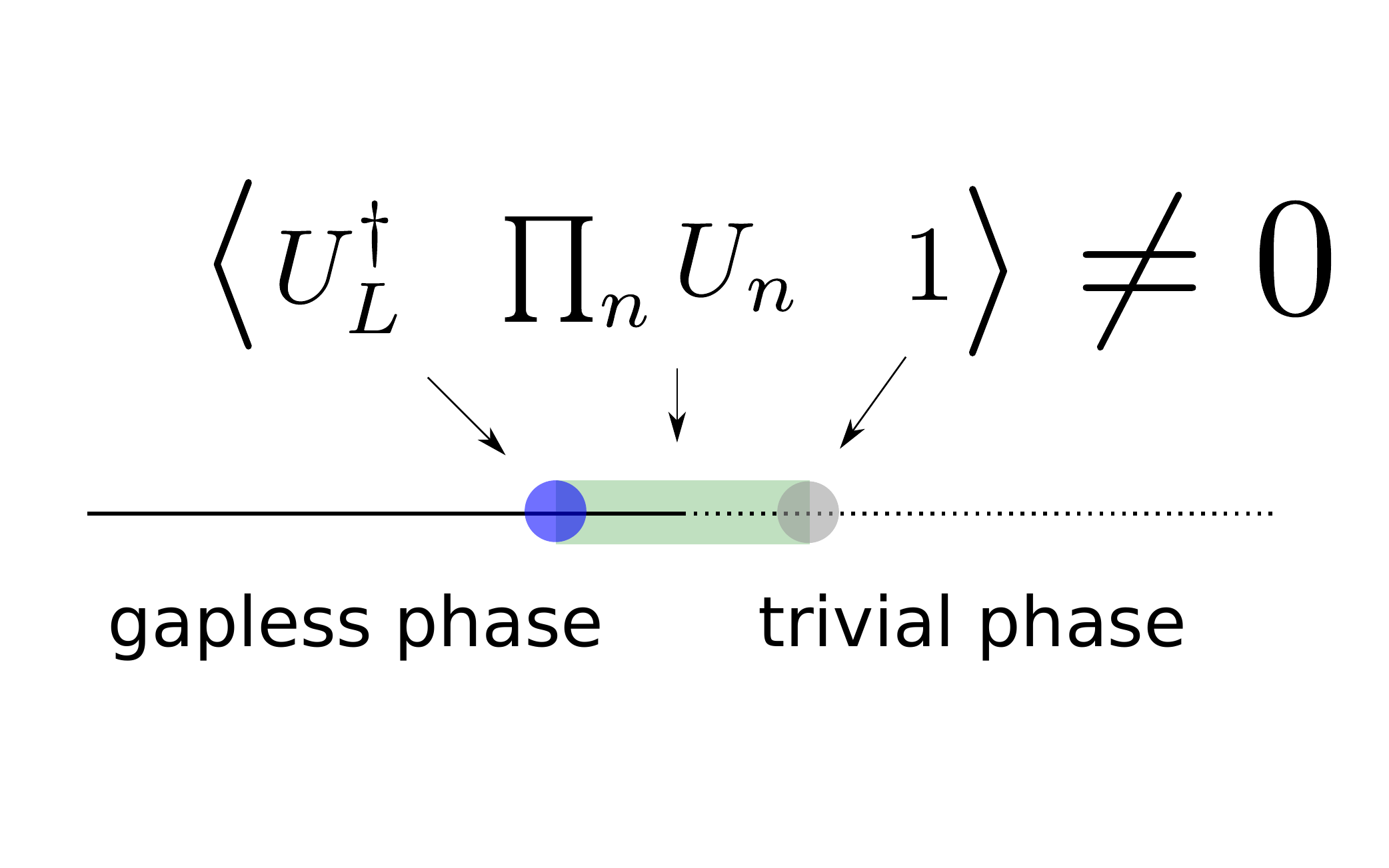}
    \caption{A gapped string operator straddling an interface between the topological phase and the trivial phase has a nonzero vev. It plays the role of a symmetry breaking order parameter at the boundary, signalling the presence of edge modes.}
    \label{figstradle}
\end{figure}

Thus, in a system with an emergent anomaly, there will be a long-range ordered string of a gapped symmetry with nontrivial charge. If we consider symmetry fractionalization at the interface between our system and a trivial phase, we then see that something must happen at the boundary. Indeed, a string which crosses the interface, with one end in the topological phase and the other in the trivial phase, will have some vev, yet carry a global charge since it has different end point in each phase. This signals a kind of spontaneous symmetry breaking at the boundary. See Fig. \ref{figstradle}.

For a finite $G$, one can construct such boundaries as follows. First we break the $G_{\rm low}$ symmetry by a generic boundary perturbation to define a ``fixed" boundary condition. $G$ will permute these boundary conditions through the action of its quotient $G_{\rm low}$. We can thus restore the symmetry by taking a direct sum of each of these $|G_{\rm low}|$-many boundary conditions.

In more familiar situations, such a ``spontaneously fixed" boundary condition would be unstable since one can add boundary-condition-changing (bcc) operators to the Hamiltonian and induce a flow to a ``free" boundary condition. These bcc operators are not local bulk operators but instead are associated with the $G_{\rm low}$ twisted sectors. As we have seen however, in the presence of an emergent anomaly, at least some of these bcc operators will be charged under $G_{\rm gap}$, so we cannot perturb by them. Some symmetry will thus remain broken at the boundary.

The sense in which the symmetries are broken is that if we consider an interval with the spontaneously fixed boundary condition on either end, the edges will polarize eachother but there will be an overall degeneracy equal to the number of broken symmetry generators. In the case of a nontrivial anomaly then, there will be a nontrivial degeneracy, which we associate to edge modes. A similar analysis was presented in \cite{verresen_gapless_2019}.

Let us relate this argument to the one given in Section \ref{secgenframe}. We would like to interpret the counterterm $\alpha$ on the RHS \eqref{eqnanommatchingalt} (see also \eqref{eqnZ4counterterm}, \eqref{eqnU(1)Z2counterterm}, \eqref{eqnU1Tcounterterm}) as defining the string charges $\chi$. Indeed, consider the partition function \eqref{eqnanompartfun}:
\[Z_{\rm low}(A_{\rm low}) e^{2\pi i\int \alpha(A_{\rm low},A_{\rm gap})}.\]
We have argued that the topological term $\alpha$ is necessary for gauge invariance. When we compute torus partition functions in the twisted sector, as we did above to measure the charges of string operators, this term contributes to the phase of the partition function, and therefore helps to encode this information. In particular, since $Z_{\rm low}(A_{\rm low})$ is independent of $A_{\rm gap}$, $\alpha$ completely encodes the $G_{\rm gap}$ charges of the different twisted sectors.

To encode such a charge, $\alpha$ contains a term schematically like $q A_{\rm gap} A_{\rm low}$. This term has a reciprocal interpretation that says the $G_{\rm gap}$ string also carries charge. This charge is what is picked up in the boundary variation when we do a $G_{\rm gap}$ gauge transformation $A_{\rm gap} \mapsto A_{\rm gap} + dg$, $\delta \alpha \ni d\left(q g A_{\rm low} \right)$. To cancel this gauge variation, we can restrict $A_{\rm low}$ at the boundary so that $q A_{\rm low} \sim 0$. This amounts to the symmetry breaking we argued above.

For instance, with $G = U(1) \times \bZ_2$ with the anomaly as in Appendix \ref{appU1timesZ2}, in light of \eqref{eqnU(1)Z2counterterm} the nontrivial $G_{\rm gap}$ charges are in the $\bZ_2$ twisted sector, so we expect there is a stable boundary condition which is ``free" from the perspective of the $U(1)$ but spontaneously fixed for the $\bZ_2$, associated with a two-fold degeneracy on an interval. The same applies to $G = U(1) \rtimes T$ of Appendix \ref{appU1timest} with $T$ spontaneously broken on an interval.

\section{Nearby Phase Diagram of the Topological Phase}\label{appnearbyphasediagram}

\begin{figure}
    \centering
    \includegraphics[width=6cm]{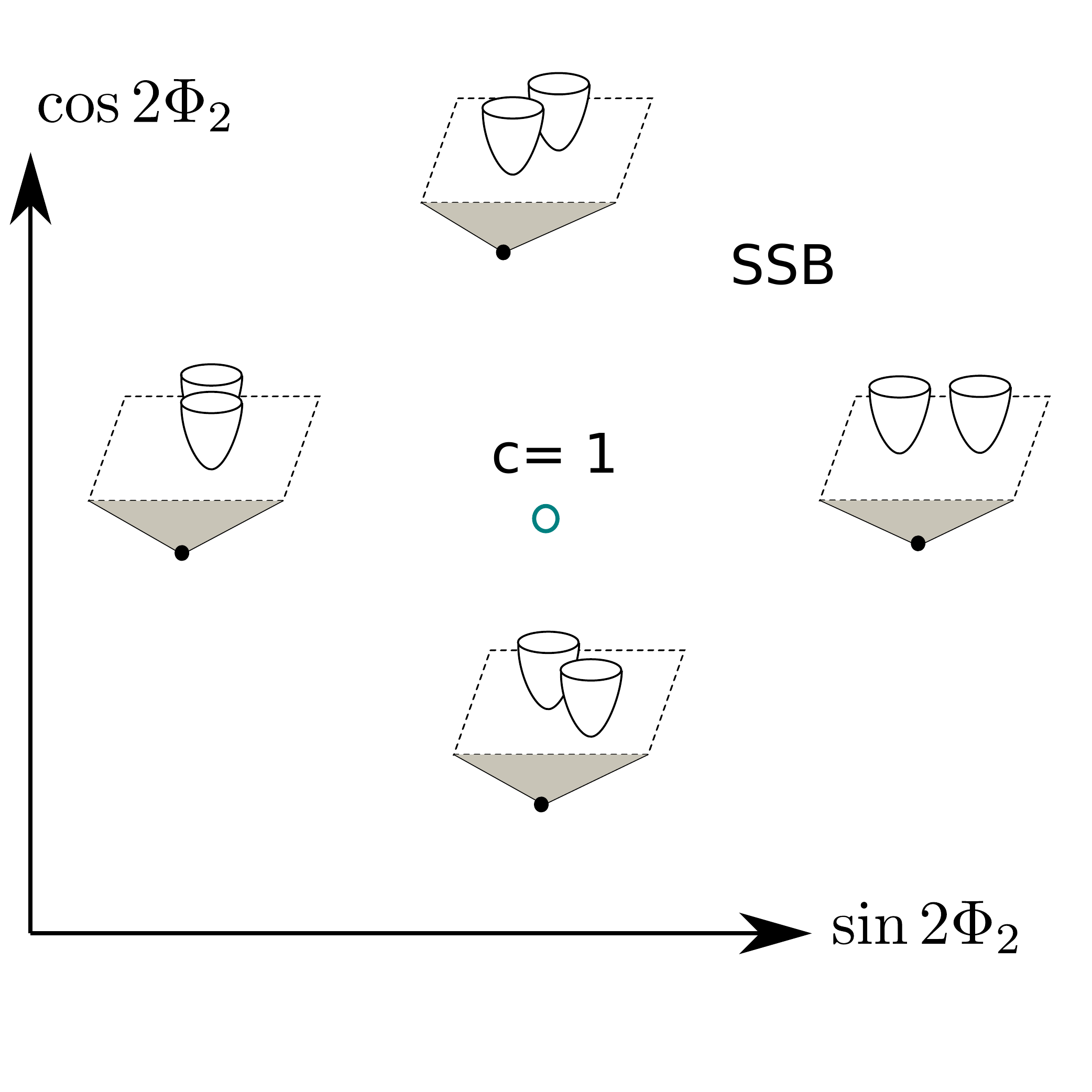}
    \caption{The gapless $\bZ_4$-SPT phase has two symmetric relevant operators, $\cos 2\Phi_2$ and $\sin 2\Phi_2$ in the region of the marginal parameter $1/8 < K_{\rm eff} < 1/2$ which describe a 2-parameter nearby phase diagram where the gapless phase is an isolated point (open blue circle) in a sea of a gapped phase where $\bZ_4$ is broken to its $\bZ_2$ subgroup. It is a diabolical point in the sense of \cite{hsin2020berry}, protected by a nontrivial vacuum crossing \cite{sharon2020global}: as one adiabatically traverses a loop encompassing the gapless point, the two SSB vacua are exchanged. A cartoon of how the mean field potential looks is illustrated at the black circles in the figure. The emergent anomaly further protects this point from deformation into an island of the trivial phase.}
    \label{fignearbyZ4}
\end{figure}

In this appendix, we discuss the nearby phase diagram of our topological gapless phase protected by $\bZ_4$ symmetry $R_x$. We use the field theory approach of Section \ref{secfieldtheory}.

% First the $R_x$-symmetric relevant operators of the $c = 2$ free fermion point are
% \[\label{eqnappperts}
% \begin{gathered}
% \cO_{CDW,1} = \cos \varphi_\uparrow + \cos \varphi_\downarrow \quad \cO_{CDW,2} = \sin \varphi_\uparrow + \sin \varphi_\downarrow \quad \cO_{x} = \cos(\phi_1/2+\phi_2/2)\cos(\theta_\uparrow - \theta_\downarrow)
% \medskip
% \medskip
% \\
% \cO_{TSC,x} = \cos(2\theta_\uparrow + 2\theta_\downarrow) \quad \cO_{TSC,y} = \sin(2\theta_\uparrow + 2\theta_\downarrow)  \quad \cO_{zz} = \cos (\varphi_\uparrow - \varphi_\downarrow)
% \end{gathered}
% \]
% The operator $\cO_{zz}$ deforms us into the topological Luttinger liquid with positive sign, or into the trivial Luttinger liquid with negative sign.

To examine the nearby phase diagram, we must first study the marginal parameters. The $c = 2$ mother theory has several such parameters. It is convenient to think of the bosonized $c = 2$ theory as a sigma model with 2-torus target and these parameters as the shape of this torus (see Appendix A of \cite{alex2020averaging} for a review). There are actually two dual tori we can think about, depending on whether we look at the coordinates $(\varphi_\uparrow,\varphi_\downarrow)$ or $(\theta_\uparrow,\theta_\downarrow)$. At the free fermion point, the torus is rectangular in $(\varphi_\uparrow,\varphi_\downarrow)$ coordinates with radii $R_\uparrow$ and $R_\downarrow$ along the $\varphi_\uparrow$ and $\varphi_\downarrow$ directions, respectively. For our models, these radii must be equal: $R_\uparrow = R_\downarrow = R = 1/\sqrt{K}$ to have $R_x$ or $T$ symmetry. These radii can be tuned together by the symmetric marginal perturbation
\[(\partial_x\varphi_\uparrow)^2 + (\partial_x\varphi_\downarrow)^2,\]
which represents a separate Luttinger interaction in each spin channel, but of the same strength. There is one more symmetric marginal perturbation:
\[\partial_x \varphi_\uparrow \partial_x \varphi_\downarrow,\]
which can be tuned by the Hubbard interaction. This corresponds to squashing the torus in the diagonal direction. We therefore write the (constant) metric as
\[g = \begin{bmatrix} R^2 && S \\ S && R^2 \end{bmatrix}.\]
The Hamiltonian term corresponding to this metric is
\[\partial_x \vec\varphi^T g \partial_x \vec\varphi.\]
The dimension of the $\varphi$ vertex operator $e^{in_\uparrow \varphi_\uparrow + i n_\downarrow \varphi_\downarrow}$ is
\[\Delta_{\vec n} = \vec n^T g^{-1} \vec n,\]
where $\vec n = (n_\uparrow, n_\downarrow)^T$. Thus $\cO_{zz} = \cos(\varphi_\uparrow - \varphi_\downarrow)$, which corresponds to a lattice vector $(1,-1)$ has dimension $2/(R^2-S)$. For it to be relevant, we need $R^2 - S > 1$.

To determine the effective radius $R_{\rm eff}$ (or effective Luttinger parameter $K_{\rm eff}$) of the topological gapless phase in terms of $R$ and $S$, we must simply see the length of the circle of minima of the potential $\cO_{zz} = \cos(\varphi_\uparrow - \varphi_\downarrow)$. This has minima along the circle $\varphi_\uparrow = \varphi_\downarrow + \pi$ embedded in the 2-torus. The length of this circle is $R_{\rm eff} = \sqrt{2R^2 + 2S}$ or
\[K_{\rm eff} = 1/(2R^2 + 2S),\]
which for the free fermion $R = 1$ and $S = 0$ is the $SU(2)$ point $K_{\rm eff} = 1/2$. At this free point, the key perturbation $\cO_{zz}$ (with dimension $2/(R^2-S)$) is marginal. To make it relevant, we can either decrease $S$, which results in $K_{\rm eff} > 1/2$, or we can increase $R$ (i.e. turn on a repulsive Luttinger interaction as in Appendix \ref{subappcoupledluttingers}), which results in $K_{\rm eff} < 1/2$. These represent two phenomenologically distinct regimes of the $c = 1$ topological phase. Below we will mainly focus on the region $K_{\rm eff} < 1/2$, which is realized by the Ising-Hubbard chain in Section \ref{seclattice} as observed in Appendix~\ref{appisinghub}.

% The $SU(2)$ spin symmetry was broken by our perturbation $\cO_{zz}$ (as well as the marginal perturbations required to make this relevant). The remaining $SU(2)$ symmetry after adding this perturbation is an accidental symmetry coming from the larger $SO(4)$ of the spinful free fermion, which can be thought of as the flavor symmetry of four Majorana fermions.

% Thus, there are essentially two phenomenologically distinct regions of the topological phase accessed by the free fermion point, depending on whether $K_{\rm eff}$ is slightly above or below the self-dual point $K_{\rm eff} = 1/2$.

Recall in the new variables $\Phi_2 = \varphi_\downarrow$, $\Theta_2 = \theta_\uparrow + \theta_\downarrow - \varphi_\uparrow/2 + \varphi_\downarrow/2$, we have the symmetry action
\[
R_x:\begin{cases} \Phi_2 \mapsto \Phi_2 + \langle \Phi_1 \rangle \\
\Theta_2 \mapsto \Theta_2 + \langle \Phi_1 \rangle,
\end{cases}
\]
where in the topological phase we have $\langle \Phi_1 \rangle = \pi$. In this case, the most relevant symmetric operators are $\cos 2\Theta_2$, $\sin 2\Theta_2$, $\cos 2\Phi_2$, $\sin 2 \Phi_2$ (there are also $\cos \Theta_2 \cos \Phi_2$ etc. but these operators cannot open a gap). The first two are relevant for $K_{\rm eff} > 1/2$ and the second two for $K_{\rm eff} < 1/2$. As long as $K_{\rm eff}$ is within the window $(1/8,2)$, these are no other symmetric relevant operators. For any combination of each pair, the system flows to an SSB state where $\bZ_4$ is broken to $\bZ_2^F$. The nearby phase diagram for $1/8 < K_{\rm eff} < 1/2$ is drawn in Fig. \ref{fignearbyZ4}.

% In this phase diagram, the relevant operators $\cos 2\Theta_2$ and $\sin 2\Theta_2$ are superconducting pairing terms, preserving a $\bZ_2$ subgroup of the particle number $U(1)$. The reason the superconducting pairing leads to spontaneous $R_x$ symmetry breaking is that domain walls in the antiferromagnetic order are bound to holes in the topological phase. Once these holes form bound states, the domain walls can no longer fluctuate and the antiferromagnetic order is restored.

In the region $K_{\rm eff} < 1/2$, at generic filling, translation symmetry stabilizes the gapless phase. In the region $K_{\rm eff} > 1/2$, it is stabilized by particle number $U(1)$ symmetry. In this latter case the $\bZ_2$ symmetry generated by the product of an order four element of $U(1)$ and $R_x$ forms an SPT phase with $\bZ_2^F$. To have a stable gapless phase but still be in a symmetry class with no gapped SPTs, we can take our symmetry group to be $\bZ_{12} = \bZ_4 \times \bZ_3$ generated by $R_x$ and the $\bZ_6$ subgroup of $U(1)$.

\section{Field Theory for Other Symmetry Classes}\label{appothersymmetryclasses}

In Section \ref{secfieldtheory} we studied an intrinsically gapless $\bZ_4$-SPT phase in proximity to a free spinful fermion. In this appendix, we will show that the same field theory captures topological phases in the symmetry classes $U(1) \times \bZ_2$ and $U(1) \rtimes T$ as well.

\subsection{Other Fermionic Classes}

\subsubsection{Symmetries}

Besides our $\bZ_4$ symmetry $R_x$, we are also interested in time reversal and particle number conservation. These act on the bosonic variables as follows:
\[\label{eqnsymmactionsapp}
R_x: \begin{cases}
\varphi_s \mapsto \varphi_{-s} \\
\theta_\uparrow \mapsto \theta_\downarrow +\pi/2 \\
\theta_\downarrow \mapsto \theta_\uparrow-\pi/2 \\
U_\uparrow \mapsto U_\downarrow \\
U_\downarrow \mapsto - U_\uparrow,
\end{cases} \qquad T: \begin{cases}
\varphi_{s} \mapsto \varphi_{-s} \\
\theta_s \mapsto -\theta_{-s}\\
U_\uparrow \mapsto U_\downarrow \\
U_\downarrow \mapsto - U_\uparrow,
\end{cases} \qquad U(1): \begin{cases}
\varphi_s \mapsto \varphi_s \\
\theta_s \mapsto \theta_s + \alpha \\
U_s \mapsto U_s.
\end{cases}
\]
We have $T^2 = (-1)^F$ and fermion parity $(-1)^F$ is a $U(1)$ rotation with $\alpha = \pi$. For convenience, we also collect
\[e^{i\alpha S_z}: \begin{cases}
\varphi_s \mapsto \varphi_s \\
\theta_\uparrow \mapsto \theta_\uparrow + \alpha/2 \\
\theta_\downarrow \mapsto \theta_\downarrow - \alpha/2  \\
U_s \mapsto U_s
\end{cases} \qquad 
{\rm Trans}: \begin{cases}
\varphi_s \mapsto \varphi_s + k_R - k_L \\
\theta_s \mapsto \theta_s + k_R/2 + k_L/2  \\
U_s \mapsto U_s.
\end{cases}\]
We see the operator $\cO_{zz} = \cos(\varphi_\uparrow-\varphi_\downarrow)$ which tunes between the trivial and topological gapless phases remains symmetric under $U(1)$, $T$, and translation.

\subsubsection{The Topological Phase}

The variables of section \ref{secfieldtheory} can be derived by first performing the following $SL(2,\bZ)$ change of variables on the bosonic variables:
\[\label{eqnchangevarsapp}\begin{gathered}
\Phi_1 = \varphi_\uparrow - \varphi_\downarrow \\
\Phi_2 = \varphi_\downarrow\end{gathered}\qquad \begin{gathered}
\Theta_1 = \theta_\uparrow \\
\Theta_2' = \theta_\uparrow + \theta_\downarrow,
\end{gathered}\]
and then defining the gauge-invariant $\Theta_2 = \Theta_2' +\Phi_1/2$, which is conjugate to $\Phi_2$. After turning on the perturbation $\cO_{zz}$, we find the effective symmetry action
\[\begin{gathered}
    T:\begin{cases} \Phi_2 \mapsto \Phi_2 + \langle \Phi_1 \rangle \\
\Theta_2 \mapsto -\Theta_2 \end{cases} \\
U(1): \Theta_2 \mapsto \Theta_2 + 2\alpha \\
R_x:\begin{cases} \Phi_2 \mapsto \Phi_2 + \langle \Phi_1 \rangle \\
\Theta_2 \mapsto \Theta_2 + \langle \Phi_1 \rangle \end{cases} \\
{\rm Trans}: \begin{cases}
\Phi_2 \mapsto \Phi_2 + k_R - k_L \\
\Theta_2 \mapsto \Theta_2 + k_R + k_L.
\end{cases}
\end{gathered}
\]
We see in the topological phase with $\langle \Phi_1 \rangle = \pi$, the effective $U(1) \rtimes T$ symmetry has an emergent anomaly that matches the edge of a 2d bosonic topological insulator \cite{Lu_2012} (cf. Appendix \ref{appU1timest}). In this symmetry class the topological phase has no symmetric relevant operators, and defines a stable gapless phase.

\begin{figure}
    \centering
    \includegraphics[width=6cm]{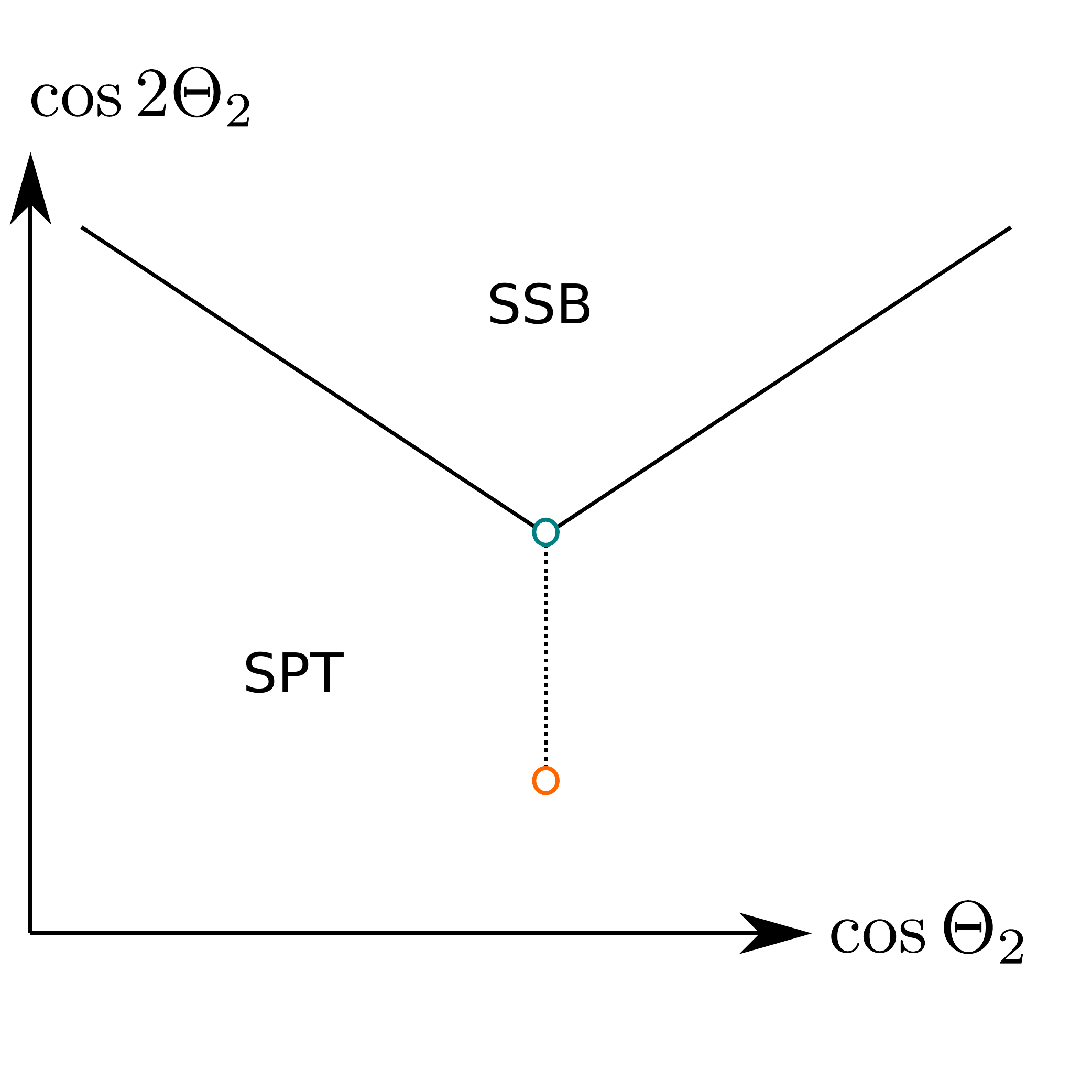}
    \caption{The phase diagram near the topological gapless phase (blue circle) in the symmetry class $T^2 = (-1)^F$, for which there is no emergent anomaly, in a region of parameter space with effective marginal parameter $1/2 < K_{\rm eff} < 2$. In the SSB phase, $T$ is broken to fermion parity, while in the nondegenerate gapped phase, which is the SPT phase for $T$, there is a spurious first order line, explained that in the effective $T^2 = 1$ symmetry of the gapless phase (where fermions are gapped), the two nearby gapped phases look like distinct SPTs. We expect this first order line to end in an Ising critical point where the fermion gap closes. One way to account for the topological nontriviality of this critical point is to note that it has no nearby trivial phase. See \cite{verresen2020topology} for a discussion of this philosophy and how it results in edge modes. The black lines are (SPT-twisted) Ising phase transitions controlling the spontaneous symmetry breaking of $T$, and are also expected to have edge modes \cite{verresen_gapless_2019}. The edge mode at the multicritical point where the two lines meet can be considered the continuation of the edge modes from either side. $R_x$ acts by reflecting this phase diagram over the first order line, i.e. $\cos \Theta_2 \mapsto - \cos \Theta_2$, and is spontaneously broken along it. In the symmetry class $\bZ_4^{R_x} \times \bZ_2^{R_x T}$ generated by $R_x$ and $T$ therefore this critical point is like a deconfined quantum critical point between two different ordered phases with an emergent anomaly for $\bZ_2^{R_x} \times \bZ_2^T$.}
    \label{figTphasediagram}
\end{figure}

With just $T$ symmetry, and $1/2 < K_{\rm eff} < 2$ (see Appendix \ref{appnearbyphasediagram}) there are two symmetric relevant operators, $\cos \Theta_2$ and $\cos 2\Theta_2$. From the perspective of the low energy theory, as we tune $\cos \Theta_2$, the gapless topological phase looks like a codimension 2 critical point between a trivial phase and the bosonic $T$ SPT. However, since the microscopic symmetry has $T^2 = (-1)^F$, these phases are actually equivalent (and actually realize a different nontrivial SPT, see below). Thus from this point of view our topological phase is reminiscent of an ``unnecessary" (multi-)critical point in the language of \cite{Bi_2019}. See Fig. \ref{figTphasediagram}.

\begin{figure}
    \centering
    \includegraphics[width=6cm]{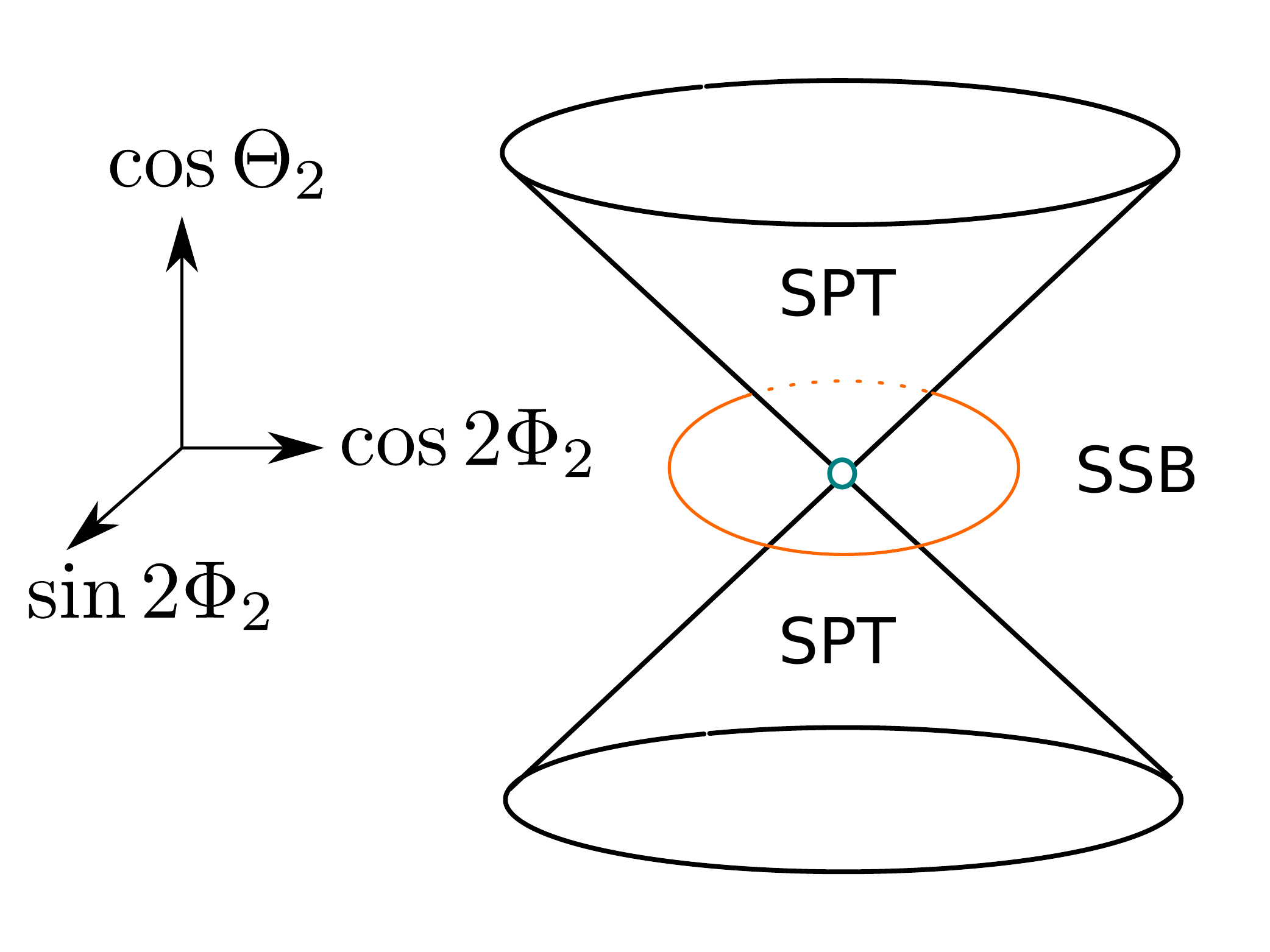}
    \caption{The phase diagram near the topological gapless phase (blue circle) in the symmetry class $T^2 = (-1)^F$ again, but this time in a region of parameter space with effective marginal parameter $1/8 < K_{\rm eff} < 1/2$. In the SSB phase, $T$ is broken to fermion parity, and both SPT phases are the unique $\bZ_4^T$ SPT phase. Along the orange circle, we have vacuum crossing as in Fig. \ref{fignearbyZ4}, which is reproduced along the plane where we turn off the $\cos \Theta_2$ perturbation, which is odd under $R_x$.}
    \label{figTphasediagram2}
\end{figure}

We note that the $T$ symmetry class is enough to project the edge modes of the topological phase. The trick is to realize that if we include the gapped sector, then the full $T^2 = (-1)^F$ symmetry is in an SPT phase (such phases are classified by a $\bZ_2$ invariant). This follows from our calculation in Appendix \ref{appU1timest} as well as the form of the string operator from Section \ref{seclattice}, namely $S^z_m (\prod_{m < k < n} P_k) S^z_n$, since $S^z$ is odd under $T$.

% We can also see the SPT phase from the $c = 2$ point. If we first gap $\Theta_2$ with $\cos \Theta_2$, then the effective symmetry on the remaining gapless fields is
% \[T: \begin{cases}\Phi_1 \mapsto -\Phi_1 \\ \Theta_1 \mapsto \Theta_1 + \pi/2 \end{cases}.\]
% Now we see that our perturbation $\cO_{zz} = \cos \Phi_1$ realizes this $c = 1$ as a transition from the trivial gapped phase, with $\langle \Phi_1\rangle = 0$ and the $T$-SPT with $\langle \Phi_1 \rangle = \pi$. This theory has no other symmetric relevant operators, so it is a direct SPT transition.

We can also define a $\bZ_2$ symmetry $Q$ by $R_x$ times the $U(1)$ rotation with $\alpha = \pi/2$. The effective $U(1) \times \bZ_2^Q$ symmetry of the topological gapless phase has the emergent anomaly of Appendix \ref{appU1timesZ2}, where $Q: \Phi_2 \mapsto \Phi_2 + \pi$ and $U(1):\Theta_2 \mapsto \Theta_2 + 2\alpha$. In the symmetry class we see there no symmetric relevant operators, and therefore our topological gapless phase is stable.

As in the case of $T$ symmetry, if we break the $U(1)$ down to fermion parity, the edge mode of the topological phase is protected by a $\bZ_2 \times \bZ_2$ SPT phase between $Q$ and $(-1)^F$, the latter of which acts only in the gapped sector. This follows from our calculation in Appendix \ref{appU1timesZ2} as well as the form of the string operator  $S^z_m (\prod_{m < k < n} P_k) S^z_n$, since $S^z$ is $Q$-odd.

\subsection{A Field Theory For Bosons}\label{subappbosons}

This field theory can be easily adapted to describe intrinsically gapless phases of bosons as well. For symmetries, we take
\[
U: \begin{cases}
\varphi_{1,2} \mapsto \varphi_{2,1} \\
\theta_{1,2} \mapsto \theta_{2,1} + \pi/2,
\end{cases} \qquad T: \begin{cases}
\varphi_{1,2} \mapsto \varphi_{2,1} \\
\theta_\uparrow \mapsto -\theta_\downarrow + \pi/2 \\
\theta_\downarrow \mapsto -\theta_\uparrow - \pi/2
\end{cases} \qquad U(1): \begin{cases}
\varphi_{1,2} \mapsto \varphi_{1,2} \\
\theta_{1,2} \mapsto \theta_{1,2} + \alpha.
\end{cases}
\]
Our perturbation into the topological phase is $\cos(\varphi_\uparrow - \varphi_\downarrow)$, which we analyze by passing to the variables
\begin{equation}
\begin{array}{ll}
\Phi_1 = \varphi_\uparrow - \varphi_\downarrow   &\qquad \Theta_1 = \theta_\uparrow \\
\Phi_2 = \varphi_\downarrow &\qquad \Theta_2 = \theta_\uparrow + \theta_\downarrow.
\end{array}
\end{equation}
The effective symmetry action is
\[\begin{gathered}
    T:\begin{cases} \Phi_2 \mapsto \Phi_2 + \langle \Phi_1 \rangle \\
\Theta_2 \mapsto -\Theta_2 \end{cases} \\
U(1): \Theta_2 \mapsto \Theta_2 + 2\alpha \\
U:\begin{cases} \Phi_2 \mapsto \Phi_2 + \langle \Phi_1 \rangle \\
\Theta_2 \mapsto \Theta_2 + \pi.
\end{cases}
\end{gathered}
\]
We recognize the $\bZ_2^U$ action as the Levin-Gu/CXZ anomaly. $U(1) \rtimes T$ matches the edge of a 2d bosonic topological insulator \cite{Lu_2012}, and $\bZ_2^U \times U(1)$ a kind of bosonic quantum spin Hall SPT.

\section{Note on Bosonization}\label{appbosonization}

Bosonization takes many forms. Perhaps the most familiar is the Jordan-Wigner transformation, which abstractly describes a fermionic theory as a bosonic theory (``the bosonization") with a $\bZ_2$ global symmetry. The fermionic theory is reconstructed by gauging this global symmetry in the presence of a modified Gauss law, such that we project out $\bZ_2$ charged local operators, but keep $\bZ_2$ charged string operators (these become the local fermions) and project out $\bZ_2$ neutral strings. See for instance \cite{Chen_2018,chen2019exact}. This fermionization transformation can also be thought of as gauging the $\bZ_2$ symmetry in the presence of a special discrete torsion which couples the gauge field to the spin structure \cite{Gaiotto_2016,thorngren2018anomalies}. It can be confusing but it is useful to be clear about such issues when questions of global topological properties are important. An advantage of JW bosonization is that all fermion operators anticommute by virtue of being charged endpoints of the same string.

The bosonization used in Section \ref{secfieldtheory} is a little bit different from the familiar Jordan-Wigner transformation. Instead, the fermionization  transformation involves gauging a $\bZ_2 \times \bZ_2$ global symmetry. For the $c = 2$ theory, these symmetries are
\[U_B:\theta_\uparrow \mapsto \theta_\uparrow + \pi \qquad U_C:\theta_\downarrow \mapsto \theta_\downarrow + \pi.\]
We project out all local operators which are charged under them, but we keep certain charged twist operators. From the three twisted sectors, corresponding to $U_A$, $U_B$, and $U_A U_B$, respectively, the operators we keep are those of the form
\[e^{i \varphi_\uparrow/2 + i \theta_\uparrow} \cO \qquad e^{i \varphi_\downarrow/2 + i \theta_\downarrow} \cO' \qquad e^{i \varphi_\uparrow/2 + i \theta_\uparrow} e^{i \varphi_\downarrow/2 + i \theta_\downarrow} \cO''\]
where $\cO$ etc are gauge invariant local operators. The operators $e^{i \varphi_j/2 + i \theta_j}$ are the two fermions. This rule may be stated that an operator in the $U_A^m U_B^n$ twisted sector is local iff it has $U_A$ charge $(-1)^m$ and $U_B$ charge $(-1)^n$.

Fermion parity in this theory is the diagonal magnetic symmetry $\varphi_j \mapsto \varphi_j + 2\pi$, which is equivalent to $\theta_j \mapsto \theta_j + \pi$ by the selection rules. The selection rules mean that certain innocent-looking operators such as $\cos(\theta_\uparrow + \theta_\downarrow)$, which appear to be even under fermion parity, are actually not local because they don't obey the gauge constraints. The bosonic operator $\cos(\varphi_\uparrow/2+\varphi_\downarrow/2)\cos(\theta_\uparrow + \theta_\downarrow)$ however, is local.

In the language of \cite{Gaiotto_2015,thorngren2018anomalies}, we are gauging the $\bZ_2 \times \bZ_2$ global symmetry of the product of the two bosonized theories, in the presence of the discrete torsion $(-1)^{Q_\eta(B) + Q_\eta(C)}$, where $B$ and $C$ are $\bZ_2$ gauge fields, $\eta$ is the spin structure, and $Q_\eta$ is the quadratic form associated to it.

\section{Discrete Gauge Field Primer}\label{appgaugefield}

Here we collect some basic definitions and facts about discrete gauge fields. A good reference for the general topological notions is \cite{hatcher2002algebraic}.

\subsection{Discrete Gauge Fields}

Let $G$ be a finite group, possibly nonabelian. To describe a $G$ gauge field on a space $X$, we choose a triangulation of $X$. A flat $G$ gauge field is a 1-cocycle $A \in Z^1(X,G)$ meaning to each oriented edge $(ij)$ (``$i$ to $j$") it assigns an element $A(ij) \in G$ such that $A(ji) = A(ij)^{-1}$ and for every triangle $(ijk)$ with ordered vertices (``$i$ to $j$ to $k$") it satisfies the cocycle condition
\[\label{eqncocyclecond}A(ij)A(jk) = A(ik),\]
which says that the magnetic flux through this triangle vanishes. A gauge transformation is parametrized by a 0-cochain $g \in C^0(X,G)$ meaning that to each vertex $i$ it assigns an element $g(i) \in G$. It acts on $A$ by
\[A(ij) \mapsto A^g(ij) = g(i)^{-1} A(ij) g(j).\]
Since we have written $A$ multiplicatively, the Wilson loop along some oriented cycle $\gamma = (i_1 i_2) + (i_2 i_3) + \cdots + (i_n i_1)$ is
\[W_R(\gamma,A) = {\rm Tr}\left(\prod_j \rho(A(i_j i_{j+1}))\right),\]
where $\rho$ is a representation of $G$.

We will be interested in studying the Euclidean partition function on an arbitrary spacetime $X$ equipped with a flat $G$ gauge field. It is so far unclear how to do this for a general lattice model, especially when $X$ has interesting topology. In field theory, it is more-or-less clear however what one needs to do. That is, we think about the gauge field, defined on a triangulation, in terms of a Poincar\'e dual network of topological defects \cite{Gaiotto_2015}. That is, in the cell structure dual to the triangulation of $X$, there are hypersurface associated to each edge $(ij)$. We choose boundary conditions for the fields across these hypersurfaces such that when a field crosses the hypersurface along the direction $i \to j$, it satisfies the gluing condition $\phi_L \cdot  A(ij) = \phi_R$, where $\phi_L \cdot A(ij)$ is the action of the global symmetry. The cocycle condition says that at a codimension 2 junction of hypersurfaces, the boundary condition is well-posed. Otherwise there would be a singularity in the fields and we would have a $G$ flux. The triangulation considered above should be considered on a length scale where the continuum field theory reigns.

For more general ``quantum symmetries", whose action cannot be seen on the fields, we associate a topological defect to each of these hypersurfaces, defined by the group element $A(ij)$. If the topological defect associated to $g \in G$ is placed along a space-like slice of the theory, then in the partition function we apply the global $g$ symmetry at that moment in time. If the topological defect is placed along some hypersurface with one coordinate in the time direction and the rest along the spatial directions, then where the spatial slice intersects it we are considering a $g$-twisted sector of the Hilbert space.

Gauge transformations allow us to freely move and recombine this network of topological defects without changing the value of the partition function. In the case of an anomaly however, the phase of the partition function will change under certain gauge transformations. In 1+1D, the basic recombination of the network of topological line defects is the ``$F$-move" or ``crossing relation". The anomaly in this case is captured by the $F$-symbols of the 2+1D Dijkgraaf-Witten theory obtained by gauging the global symmetry of the associated bulk SPT. See \cite{barkeshli_symmetry_2014} for more details.

% (TODO: clean this sentence) In order to couple a continuum field theory to a discrete gauge field on $d+1$-dimensional spacetime, we need to associate the global $G$ symmetry with $d$-dimensional invertible topological defects labelled by group elements which fuse according to the group multiplication rules \cite{Gaiotto_2015}. The triangulation is thus defined on some very large length scale and serves only to specify the combinatorics of a network of such topological defects, such that an edge $(ij)$ labelled with a group element $A(ij) \in G$ corresponds to an $A(ij)$ topological defect (which is codimension 1 in spacetime) intersecting that edge at a point. The cocycle condition on triangles tells us that the fusion rules of these topological defects are satisfied at codimension 2 junctions, and a gauge transformation is a recombination of the network.

% For 1+1D systems, these topological defects are line operators, and all recombinations of the network are expressible as a sequence of ``$F$-moves". Anomalies are associated with phase factors obtained under such moves \cite{chang_topological_2019,Lin_2019}. We will see below how to derive these phase factors from group cohomology. Matching the anomalous phase factors with the group cohomology class allows us to classify the anomaly of the 1d theory.

\subsection{Simplicial Cohomology}

We will need some basic constructions from simplicial cohomology. First, a $k$-simplex is determined by its $k+1$ vertices. If these vertices are ordered $v_0 <  \cdots < v_k$ we write $(v_0 \cdots v_k)$ to denote the corresponding $k$-simplex with this vertex ordering. When we are talking about a single $k$-simplex, we will just write it with the shorthand $(0 \cdots k)$.

We define the group of $k$-chains $C_k(X,\bZ)$ on a triangulated space $X$ to be integer combinations of $k$-simplices with ordered vertices, modulo reordering of the vertices: $( \sigma(0) \cdots \sigma(k)) = (-1)^\sigma (0 \cdots k)$.

Let $M$ be an abelian group. We define a $k$-cochain $\alpha \in C^k(X,M)$ to assign an element $\alpha(0 \cdots k) \in M$ to every $k$-simplex $(0 \cdots k)$ with ordered vertices such that $\alpha( \sigma(0) \cdots \sigma(k)) = (-1)^\sigma \alpha(0 \cdots k)$ for any permutation $\sigma \in S_{k+1}$ of the vertices with sign $(-1)^\sigma$. From now on we just refer to $(1 \cdots k)$ as a $k$-simplex.

For $\Gamma \in C_k(X,\bZ)$, $\alpha \in C^k(X,M)$ integral
\[\int_\Gamma \alpha \in M\]
is defined to be the sum of $\alpha$ evaluated on these $k$-simplices, weighted by those integers. Thus by definition
\[\int_{(0 \cdots k)} \alpha = \alpha(0 \cdots k),\]
so $\alpha$ is determined by its integrals. In this sense, $\alpha$ is like a discrete analog of a differential $k$-form.

Taking the boundary of a $k$-simplex with ordered vertices, and passing that ordering onto its boundary $k-1$-simplices produces a $k-1$ chain, and extending that function by linearity defines the boundary map $\partial: C_k(X,\bZ) \to C_{k-1}(X,\bZ)$. We define the group of $k$-cycles as $Z_k(X,\bZ) := {\rm ker}(d)$ and the group of $k$-boundaries as $B_k(X,\bZ) := {\rm im}(d)$. We have $\partial^2 = 0$, so $B_k(X,\bZ) \subset Z_k(X,\bZ)$. The homology is defined by $H_k(X,\bZ) := Z_k(X,\bZ)/B_k(X,\bZ)$.

We define the differential $d:C^k(X,M) \to C^{k+1}(X,M)$ by Stokes' theorem
\[\int_{\Gamma} d\alpha = \int_{\partial \Gamma} \alpha,\]
which translates to
\[\label{eqndifferential}(d\alpha)(0\cdots k+1) = \sum_{j=0}^{k+1} (-1)^j\alpha(0 \cdots \hat j \cdots k+1)\]
\[= \alpha(1 2 3 \cdots k+1) - \alpha(0 2 3 \cdots k+1) + \alpha(0 1 3 \cdots k+1) - \cdots + (-1)^{k+1} \alpha(0 \cdots k),\]
where $\hat j$ means $j$ is excluded from $(0 \cdots j \cdots k+1)$. We define the group of $k$-cocycles as $Z^k(X,M) := {\rm ker}(d) \subset C^k(X,M)$ and the group of $k$-coboundaries as $B^k(X,M) := {\rm im}(d) \subset C^k(X,M)$. We find $d^2 = 0$ so $B^k(X,M) \subset Z^k(X,M)$. The cohomology group is defined by $H^k(X,M) := Z^k(X,M)/B^k(X,M)$. 

We can connect this to two physical objects: gauge fields and partition functions.

Firstly, when $G$ is abelian, $A \in C^1(X,G)$, and the cocycle condition \eqref{eqncocyclecond} for a $G$ gauge field can be rewritten
\[dA = 0,\]
and a gauge transformation acts by
\[A \mapsto A^g = A + dg,\]
with $g \in C^0(X,G)$ so gauge equivalence classes of flat $G$ gauge fields on $X$ are classified by $H^1(X,G)$. (This in fact also holds for nonabelian $G$, although in that case this construction does not generalize to $H^k(X,G)$.)

Secondly, if $X$ is a spacetime of dimension $k = d+1$, then $H^k(X,U(1))$ (where we identify $U(1) \cong \mathbb R / \mathbb Z$) corresponds to topological actions. More precisely, any $\omega \in H^k(X,U(1))$ defines a partition function $Z = e^{2\pi i \int_X \omega}$ for any closed manifold $X$. The fact that $d \omega=0$ physically implies cobordism invariance of $Z$.

To connect these two, i.e., the gauge field $A$ and the action $\omega$, we turn to group cohomology.

\subsection{Group Cohomology and Topological Terms}

There is a useful space called the classifying space, denoted $BG$, which has the property that a flat $G$ gauge field on $X$ is the same as a map $X \to BG$. A construction can be found in \cite{hatcher2002algebraic}. See also \cite{brown2012cohomology}. The group cohomology is defined to be the cohomology of this space: $H^k(BG,M)$.
% There is a useful space called the classifying space, denoted $BG$, which for a discrete group $G$ can be constructed as follows. It has a single vertex $\circ$ and an edge for each group element, indicated
% \[\circ \xrightarrow{g} \circ.\]
% Then it has a triangle for each pair of group elements $g_1$ $g_2$, indicated
% \[\circ \xrightarrow{g_1} \circ \xrightarrow{g_2} \circ\]
% with boundary edges
% \[\partial(\circ \xrightarrow{g_1} \circ \xrightarrow{g_2} \circ) = (\circ \xrightarrow{g_2} \circ) - (\circ \xrightarrow{g_1g_2} \circ) + (\circ \xrightarrow{g_1} \circ),\]
% as so on for $k$-simplices of all $k$. By construction, $BG$ has the property that a map $X \to BG$ which sends vertices to vertices, edges to edges, and so on, is equivalent to a flat $G$ gauge field on $X$.
An element of $H^k(BG,M)$ can be represented by an $M$-valued group $k$-cocycle which is a function $\Omega:G^k \to M$ that satisfies the group $k$-cocyle equation
\[\Omega(g_2, g_3, g_4 \ldots, g_{k+1}) - \Omega(g_1 g_2, g_3, g_4, \ldots, g_{k+1}) + \Omega(g_1,g_2 g_3, g_4, \ldots, g_{k+1}) - \cdots + (-1)^{k+1}\Omega(g_1, \ldots, g_k) = 0.\]
One can use such a $k$-cocycle to define a map (not necessarily a group homomorphism)
\[\omega:Z^1(X,G) \to Z^k(X,M)\]
which satisfies the $k$-cocycle equation
\[\label{eqnomegacocycle}d\omega(A) = 0\]
and the gauge-invariance equation
\[\label{eqnomegainvariance}\omega(A^g) = \omega(A) + d\omega_1(A,g)\]
for some $\omega_1(A,g) \in C^{k-1}(X,M)$. To do so, we define
\begin{equation}
\omega(0 \cdots k) = \Omega(A(01),A(12),\ldots,A(k-1,k)). \label{def:omega}
\end{equation}
(In terms of $A$ considered as a map $\mathcal{A}:X \to BG$, $\omega(A) = \mathcal{A}^* \Omega$, where $\mathcal{A}^*$ is the pullback \cite{hatcher2002algebraic}.) $\Omega$ can be reconstructed from $\omega$ by evaluating $\omega$ on a $k$-simplex.

For $M = U(1) = \mathbb{R/Z}$, $k = d+1$ the dimension of spacetime, such a cocycle defines a topological action for the $G$ gauge field \cite{dijkgraafwitten}:
\[Z_{\rm top}(X,A) = e^{2\pi i \int_X \omega(A)}.\]
We think of this partition function as arising from a $d$-space-dimensional SPT coupled to the background gauge field $A$. It turns out that in a certain sense gauge invariance \eqref{eqnomegainvariance} is equivalent to the cocycle equation \eqref{eqnomegacocycle} (see below), and moreover shifts $\omega(A) \mapsto \omega(A) + d\chi(A)$ do not change the partition function on closed spacetime. Thus, the gauge invariant partition functions of this form on closed spacetimes are classified by $H^{d+1}(BG,U(1))$. If we assume that these partition functions capture all the topological features of the $G$ symmetry of the underlying model, we reproduce the group cohomology classification of SPTs from \cite{Chen_2013}. See also \cite{kapustin2014bosonic,kapustin2014symmetry} for generalizations.

The equivalence between gauge invariance \eqref{eqnomegainvariance} and the cocycle equation \eqref{eqnomegacocycle} is very important. It underlies our anomaly vanishing equation as well as the calculations of anomalies in \cite{kapustin2014anomalies}. A mathematical proof can be found in \cite{Thorngrenthesis}, which we reproduce here.

The proof relies on being able to study $\omega$ on different $X$, so that $\omega$ is not allowed to depend on precise details of $X$. Intuitively it should be a function of the gauge field only. The precise condition is that $\omega$ is a natural transformation of functors from $Z^1(-,G)$ to $Z^{d+1}(-,U(1))$. 

To show that gauge invariance \eqref{eqnomegainvariance} implies the cocycle equation \eqref{eqnomegacocycle}, we consider $X = \Delta^{d+2}$ a $d+2$ simplex with flat gauge field $A$ and
\[\int_{\Delta^{d+2}} d\omega(A) = \int_{\partial \Delta^{d+2}} \omega(A).\]
Because $\Delta^{d+2}$ is contractible, $A = 1^g$ for some $g$ (i.e. it is a gauge transformation of the trivial gauge field $A = 1$, written in the multiplcative notation). Using \eqref{eqnomegainvariance}, we then complete the proof
\[\int_{\Delta^{d+2}} d\omega(A) =\int_{\partial \Delta^{d+2}} \omega(A) = \int_{\partial \Delta^{d+2}} \omega(1) - d\omega_1(1,g) = 0.\]

To show the other direction, for any $A \in Z^1(X,G)$, $g \in C^0(X,G)$, we can define a flat gauge field $\hat A$ on $X \times [0,1]$ such that $A|_{X \times 0} = A$ and $A|_{X \times 1} = A^g$. Using the cocycle equation, we have
\[0 = \int_{X \times [0,1]} d\omega(\hat A) = \int_{\partial(X \times [0,1])} \omega(\hat A) = \int_X \omega(A^g) - \int_X \omega(A) + \int_{\partial X \times [0,1]} \omega(\hat A|_{\partial X \times [0,1]}).\]
The last term can be integrated over the interval to define
\[\omega_1(A,g) := -\int_{[0,1]} \omega(\hat A|_{\partial X \times [0,1]}).\]
Gauge invariance \eqref{eqnomegainvariance} follows.

This argument is very general, and can be easily adapted to \textit{relative} cohomology, in the form it is used to prove the anomaly vanishing equation. That is, if we have a $G_{\rm low}$ gauge field $A_{\rm low}$ on $X$ and a $G$ gauge field $A$ on $\partial X$, we are interested in invariants of the form
\[\int_X \omega(A_{\rm low}) - \int_{\partial X} \alpha(A). \]
We find that gauge invariance of such expressions is equivalent to $d\omega(A_{\rm low}) = 0$ and the anomaly vanishing equation $\omega(A_{\rm low}) = d\alpha(A)$.

\subsection{Example: $\bZ_n$}

Let us discuss the special case $G = \bZ_n$ and describe the classification of its 2+1D SPTs using the method  of topological terms.

To do so we will need one more definition. Let $R$ be a ring. We define the cup product $\cup:C^j(X,R) \times C^k(X,R) \to C^{j+k}(X,R)$ by
\[(\alpha \cup \beta)(0 \cdots j+k) = \alpha(0 \cdots j)\beta(j \cdots j+k).\]
It satisfies
\[d(\alpha \cup \beta) = (d\alpha) \cup \beta + (-1)^k \alpha \cup (d\beta),\]
and if $d\alpha = 0$ and $d\beta = 0$, then
\[\alpha \cup \beta - (-1)^{jk} \beta \cup \alpha = d(\cdots),\]
where $\cdots$ define higher ``cup-$i$ products" of cochains.

It turns out that $H^3(B\bZ_n,U(1)) = \bZ_n$. The generator of this cyclic group defines, according to Eq.\eqref{def:omega}, a map $\omega: H^1(X,\bZ_n) \to H^3(X,U(1))$ given by 
%, generated by the topological term
%\[\exp\left(2\pi i \int \omega(A)\right),\]
%where
\[\omega(A) = \frac{1}{n^2} A \cup dA.\]
Note that because the denominator is $n^2$ and not $n$, we need to lift $A$ to an integer-valued cochain to define this quantity. We can do this by expressing $A(ij) \in \bZ_n$ as an integer $\tilde A(ij) \in [0,n)$, so that $\tilde A = A$ mod $n$. We see the cocycle equation for $A$ implies $d\tilde A = n \beta$ for some integer valued 2-cocycle $\beta$. $\beta$ can be thought of as the density of $2\pi$ fluxes. Thus, the form above may be re-written
\[\omega(A) = \frac{1}{n} A \cup \beta.\]
We see that if we change the lift $\tilde A \mapsto \tilde A + n \alpha$, $\beta \mapsto \beta + d\alpha$ and so
\[\omega \mapsto \omega + \frac{1}{n} A \cup d\alpha.\]
Integrating the second term by parts we find it is
\[\delta \omega = - \frac{1}{n} dA \cup \alpha + d(\cdots)\]
which is an integer on a closed spacetime and does not contribute to the phase of the partition function.

% \begin{figure}
%     \centering
%     \includegraphics[width=15cm]{Znanomaly.pdf}
%     \caption{Two sides of the boundary of a 3-simplex, with the restriction of the $\bZ_n$ gauge field to these boundaries drawn in the dual network picture as the green curves, labelled by their group elements in $\bZ_n$. The blue arrows indicate the co-orientation of these curves, which is induced from the ordering of the vertices, meaning the curve intersecting the edge $(ij)$ is labelled $A(ij)$, where $i < j$. Orange dots are $2\pi$ fluxes, where $\beta = 1$. We see that the bulk-boundary correspondence implies that these two configurations, which are related by an $F$-move, have a boundary partition function which differs by $e^{2\pi i/n}$.}
%     \label{fig:Znanomaly}
% \end{figure}

We see
\[d\omega = \frac{1}{n^2} dA \cup dA = \beta \cup \beta \in \bZ,\]
which implies gauge invariance up to a boundary term. Indeed, if we shift $A \mapsto A + dg$, $\tilde A \mapsto \tilde A + \widetilde{dg}$, so
\[\delta \omega = \frac{1}{n}\left( dg \cup \beta + A \cup d\widetilde{dg} + dg \cup d\widetilde{dg} \right) = \frac{1}{n} d\left(g \cup \beta - A \cup \widetilde{dg} + g \cup d\widetilde{dg} \right).\]

To get a handle on what this term means for the boundary anomaly, we can consider a single 3-simplex $(0123)$ with $\tilde A(01) = 1$, $\tilde A(12) = n-1$ $\tilde A(23) = 1$. We see $n\beta(123) = \tilde A(12) + \tilde A(23) - \tilde A(13) = n-1 + 1 - 0 = n$, so
\[\omega(A)(0123) = \frac{1}{n}.\]
Thus, this configuration is in some sense the minimal configuration which yields a nontrivial contribution from the topological term. It can be related to the nontrivial $F$ symbol of 2+1D $\bZ_n$ Dijkgraaf-Witten theory \cite{dijkgraafwitten}, and thus to the $\bZ_n$ anomaly.
% One way to interpret this topological term is by viewing the 3-simplex as a ``movie", which begins on one side of the boundary and ends on the other side. This is summarized in Fig. \ref{fig:Znanomaly}. We see that to diagnose the anomaly, one must equivalently look for this nontrivial $F$ move of the topological lines in the 1+1D theory.

\end{document}